\documentclass[11pt,A4,twoside]{article}
\usepackage[hmargin=1.4in,vmargin=1.2in]{geometry} 

\usepackage{graphicx}[draft]

\usepackage{natbib}
\usepackage[nottoc,notlof,notlot]{tocbibind} %
\usepackage[deletedmarkup=xout]{changes}
\usepackage{dsfont}
\usepackage{xspace}
\usepackage{mathrsfs} 
\usepackage{stmaryrd}
\usepackage{eurosym}
\usepackage{amsthm}
\usepackage{float}

\usepackage{amssymb}
\usepackage{latexsym}
\usepackage{amsmath}
\usepackage{amsfonts}
\usepackage{yhmath}

\usepackage{tikz}

\usepackage[english]{babel}
\usepackage{url}
\usepackage{enumerate}
\usepackage{hyperref}
\usepackage{color}
\usepackage[fixlanguage]{babelbib}
\selectbiblanguage{english}
\usepackage{xcolor} 
\usepackage[normalem]{ulem}
\usepackage{subcaption}

\usepackage{array}

\usepackage{amsthm}
\newtheorem{theorem}{Theorem}[section]
\newtheorem{pro}[theorem]{Proposition} 
\newtheorem{cor}[theorem]{Corollary}
\newtheorem{lem}[theorem]{Lemma}

\theoremstyle{definition}
\newtheorem{defi}{Definition}[section]
\newtheorem{hyp}[defi]{Assumption}

\theoremstyle{remark}
\newtheorem{rem}{Remark}[section] 
\newtheorem{ex}[rem]{Example}

\newcommand{\bt}{\begin{theorem}}\newcommand{\et}{\end{theorem}}
\newcommand{\bl}{\begin{lem}}\newcommand{\el}{\end{lem}}
\newcommand{\bp}{\begin{pro}}\newcommand{\ep}{\end{pro}}
\newcommand{\bcor}{\begin{cor}}\newcommand{\ecor}{\end{cor}}
\newcommand{\bconj}{\begin{conj}}\newcommand{\econj}{\end{conj}}

\newcommand{\bd}{\begin{defi} %
}\newcommand{\ed}{\end{defi} }
\newcommand{\brem }{\begin{rem} %
}\newcommand{\erem }{\end{rem}}
\newcommand{\bcom }{\begin{com} %
}\newcommand{\ecom }{\end{com}}
\newcommand{\brems }{\begin{rem} %
}\newcommand{\erems }{\end{rem}}
\newcommand{\bex}{\begin{ex} %
}\newcommand{\eex}{\end{ex}}
\newcommand{\bexo}{\begin{opt} %
}\newcommand{\eexo}{\end{opt}}

\def\bhyp{\begin{hyp} %
}\def\ehyp{\end{hyp}}

\def\proof{\noindent \emph{\textbf{Proof.} $\, $}}

\def\finproof {\hfill $\Box$ \vskip 5 pt }\def\finproof{\rule{4pt}{6pt}}\def\finproof{\ensuremath{\square}}

\DeclareMathOperator{\esssup}{esssup}

\renewcommand{\b}[1]{#1}\renewcommand{\b}[1]{#1}\renewcommand{\b}[1]{{\color{black}#1}}

\newcommand{\cT}{\mathcal T}

\newcommand{\cY}{\mathcal Y}

\newcommand{\bQ}{\mathbb Q}

\newcommand{\bF}{\mathbb F}\renewcommand{\bF}{\mathfrak F}

\newcommand{\bE}{\mathbb E}
\newcommand{\bR}{\mathbb Q}

\newcommand{\prob}[1]{\bP\left[#1\right]}\renewcommand{\prob}[1]{\Q\left[#1\right]}
\newcommand{\probc}[2]{\bP_{#1}\left[#2\right]}\renewcommand{\probc}[2]{\Q_{#1}\left[#2\right]}
\newcommand{\esp}[1]{\bE\left[#1\right]}

\newcommand{\espc}[2]{\bE_{#1}\left[#2\right]}

\newcommand{\KVA}{\mathrm{KVA}}
\newcommand{\HVA}{\mathrm{AVA}}\renewcommand{\HVA}{\mathrm{HVA}}

\def\bel{\begin{eqnarray*}\begin{aligned}}
\def\eel{\end{aligned}\end{eqnarray*}}
\newcommand{\beqa}{\begin{eqnarray}}
\newcommand{\eeqa}{\end{eqnarray}}
\def\bal{\begin{aligned}}
\def\eal{\end{aligned}}
\def\bll#1{
\beqa\label{#1}\bal
}\def\lel{\eal\eeqa}
\newcommand{\beql}[1]{\bll{#1}}
\newcommand{\eeql}{\lel}

\def\cadlag{c\`adl\`ag\xspace}

\newcommand{\indi}[1]{\ind_{\{{#1}\}}}
\def\ind{\mathds{1}}
\def\un{\mathbf{1}}
\def\qqq{\;\;\;\;\;\;\;\;\;}
\def\qq{\;\;\;\;\;\;}

\def\ba{\textit{ba}}\def\ba{0}\def\ba{*}

\def\pal{p}\def\pal{pnl}

\def\tex{\theta}\def\tex{{\tau_e}}

\def\ts{\tau}\def\ts{{\tau_{s}}}

\def\HHW{\HHW} \def\HHW{p}

\def\WW2{W^{(2)}}\def\WW2{Z}

\def\eqdef{=}

\def\pnl{pnl\xspace}

\def\EC{R}\def\EC{\mbox{ES}}\def\EC{{\rm ES}}\def\EC{{\rm EC}}
\def\ES{\mathbb{ES}}
\def\VaR{\mathbb{V}a\mathbb{R}}

\def\HVA{{\rm HVA}}

\def\AVA{{\rm AVA}}
 
\def\sp{,\,}

\def\Pi{q}

\def\theq{q} 
\def\theQ{Q}
 
\def\cQ{\mathcal{Q}}
\def\cP{\mathcal{P}}
\def\theP{P}
\def\thep{p}
\def\thepnsb{\the\thepnsb}\def\thepnsb{\thep}

\def\Tb{\overline{T}}\def\Tb{U} \def\Tb{T}

\def\E{{\mathbb E}}

\def\Athe{K}\def\Athe{\alpha}\def\Athe{{A}}
\def\theB{\beta}\def\theB{{B}}
\def\thea{{a}}
\def\theb{{b}}
\def\texnsb{\tau_e^{nsb}}
\def\texbad{\tau^{bad}_e}

\hypersetup{
 colorlinks  = true, %
 urlcolor   = black, %
 linkcolor  = black, %
 citecolor  = black %
}
\makeatletter
\renewcommand\@makefnmark{\hbox{\@textsuperscript{\normalfont\color{blue}\@thefnmark}}}
\renewcommand\@makefntext[1]{%
 \parindent 1em\noindent
      \hb@xt@1.8em{%
        \hss\@textsuperscript{\normalfont\@thefnmark}}#1}
\makeatother

\def\hurdle{r}\def\hurdle{h}
  \def\F{\mathfrak{F}}
  
\def\mkt{mkt}\def\mkt{*}\def\mkt{}

\def\even{u}
\def\odd{v}
\def\cI{\mathcal{I}}
\def\Awo{\Og}
\def\Ano{\Og}
\def\Og{\Omega}
\def\Q{\mathbb{Q}}

\def\tell{\ell}
\def\tk{k}

\def\h{ }

\def\Td{\Tb}

\def\bad{worst}\def\bad{bad\xspace}

\def\Q{\mathbb{Q}}
\def\R{\Q}

\def\bT{\mathbb T}
\def\cL{\mathcal L}

\begin{document}

\title{%
\b{The Recalibration Conundrum:} Hedging Valuation Adjustment for Callable Claims 
}
{ \let\thefootnote\relax\footnotetext{
{This research has benefited from the support of (by alphabetical order): \textit{Chair Capital Markets Tomorrow: Modeling and Computational Issues} under the aegis of the Institut Europlace de Finance, a joint initiative of Laboratoire de Probabilit\'es, Statistique et Modélisation (LPSM) / Université Paris Cit\'e and Cr\'edit Agricole CIB; \textit{Chair Futures of Quantitative Finance}, a partnership between LPSM at Université Paris Cit\'e, CERMICS at École nationale des ponts et chauss\'ees, and BNP Paribas Global Markets.}
}}
\author{Cyril B\'en\'ezet\thanks{Université Paris-Saclay, CNRS, Univ Evry, ensIIE, Laboratoire de Mathématiques et Modélisation d'Evry, 91037, Evry-Courcouronnes, France ({\tt cyril.benezet@ensiie.fr}).}
\and
St\'ephane Cr\'epey\thanks{Université
Paris Cité, Laboratoire de Probabilités, Statistique et Modélisation (LPSM), CNRS UMR 8001 ({\tt stephane.crepey@lpsm.paris}). 
} 
\and
Dounia Essaket\thanks{Université
Paris Cité, Laboratoire de Probabilités, Statistique et Modélisation (LPSM), CNRS UMR 8001 ({\tt essaket@lpsm.paris}). The PhD grant of Dounia Essaket is funded by a Bloomberg Quant Finance Fellowship.} 
}

\date{\today\\\vspace{0.5cm}
}

\maketitle

\normalem
\begin{abstract}
\b{The dynamic hedging theory only makes sense in the setup of one given model, whereas the practice of dynamic hedging is just the opposite,
with models fleeing after the data through daily recalibration. This is quite of a quantitative finance paradox.}
In this paper we  revisit  \citet*{Burnett21} \& \citet*{Burnett21b}'s notion of hedging valuation adjustment (HVA), originally intended to deal with dynamic hedging frictions, in the direction of \b{recalibration and} model risks.  Specifically, we extend to callable assets the HVA model risk approach of \citet*{AlbaneseCrepeyBenezet22}. The classical way to deal with model risk is to reserve the differences between the valuations in reference models and in the local models used by traders. However, while traders' prices are thus corrected, their hedging strategies and their exercise decisions are still wrong, which necessitates a risk-adjusted reserve. We illustrate our approach on a stylized callable range accrual representative of huge amounts of structured products on the market. We show that a model risk reserve adjusted for the risk of wrong exercise decisions may largely exceed a basic reserve only accounting for valuation differences.
\end{abstract}

\def\keywordname{{\bfseries Keywords:}}
\def\keywords#1{\par\addvspace\baselineskip\noindent\keywordname\enspace
\ignorespaces#1}

\vspace{2mm}

\vspace{2mm}

\section{Introduction}%
\label{ss:da}

The 2008 global financial crisis triggered a shift from trade-specific pricing to netting-set CVA analytics.
For tractability reasons, the market
models used by banks for their CVA analytics are simpler than the ones that they use for individual deals. Given this coexistence of models, it is no surprise if
FRTB emphasized the issue of model risk.
Traditionally, banks manage model risk by reserving the difference between 
asset valuations in reference models and in the local models used by traders, \b{which broadly corresponds to a reserve for recalibration valuation leakages}.
However, once prices are thus corrected, hedges and exercise decisions are still wrong. Hence there is residual risk and the reserve should be risk-adjusted.

In the context of structured products,
\citet*{AlbaneseCrepeyIabichino19} introduced the notion of Darwinian model risk, 
where the Darwinian terminology refers to the embedded adverse selection of local models by traders. Namely,
when a trader wants to deal a structured product with a client, the competition for clients may lead the trader to prefer a lower quality model that outputs a price more favorable to the client \textit{(first Darwinian principle)}. But the recalibration of such a model introduces alpha leakage on the asset valuation side, which thus has to be compensated on the hedging side so that the model stands a chance to be accepted by the management of the bank \textit{(second Darwinian principle)}. However, systematic gains on the hedging side of the position
is a short-to-medium viewpoint: in the long run, the falsity of the trader's model is revealed under extreme market conditions in which the local model no longer calibrates, forcing a ``bad'' trader to a suboptimal exercise  decision or a ``not-so-bad'' trader to switch to a higher quality model, at the cost of more or less substantial losses for the bank \textit{(third Darwinian principle)}.
Risk magazine thus reported that Q4 of 2019, a \$70bn notional of range accrual had to be unwound at very large losses by the industry: cf.\ {\em Remembering the range accrual bloodbath}\footnote{https://t.ly/W9ieL.} in which banks incurred losses of ``approximately \$2.5 billion" and ``never fully recovered", or {\em How axed dividends left SocGen in a \euro 200 million hole}\footnote{https://t.ly/rEFA5.}.
\citet{AlbaneseCrepeyIabichino19} argued that Darwinian model risk was key to such structured products crises.

The notion of hedging valuation adjustment (HVA) was introduced by \citet*{Burnett21} \& \citet*{Burnett21b} to account for dynamical hedging transaction costs into prices. 
As these costs are nonlinear, they cannot be assessed for individual deals, they should be computed at the hedging set  level. This feature justifies considering these costs as cross valuation adjustment (XVAs), understood as costs linked to risks, such as counterparty, funding, and capital risks, which can only be assessed at the portfolio level \citep{Crepey25}.
On top of transaction costs, \citet*{AlbaneseCrepeyBenezet22} incorporate in the HVA the impact of model risk,
accounting for recalibration valuation leakages, by setting aside as a reserve the difference between (buying) prices in bad models and prices in good models,
but also for the risk of explosion of the trader's strategy. Moreover, they risk-adjust the model risk reserve by a KVA component.
We refer to the introduction of their paper for a more extensive discussion about the genesis of HVA and a discussion of the model risk literature. More recent works related to model risk include \citet*{silotto2024xva} and \citet*{burnett2025fundamental} for model risk within a XVA environment, \citet*{sauldubois2024first} and \citet*{fan2025quantifying} for sensitivities of martingale optimal transport problems, in the line of (\citet*{bartl2021sensitivity}, and eventually \citet*{matsumoto2024multi}, \citet*{gianfreda2024assessing} and \citet*{lazar2024measures} regarding the use of risk metrics to measure model risk in various financial settings.

But \citet*{AlbaneseCrepeyBenezet22} was only focusing on European claims or portfolios. In the case of callable assets, there is also the model risk of erroneous exercise decisions, which was in fact the main motivation and focus in  \citet*{AlbaneseCrepeyIabichino19}. In the present paper, we extend to callable assets the HVA take on model risk of \citet*{AlbaneseCrepeyBenezet22},  
thus providing mathematical and quantitative foundations to \citet*{AlbaneseCrepeyIabichino19}'s pioneering intuition.   
As an illustration of our approach, we devise an explicit example, stylized but representative \b{of a very popular and liquid structured product on the market, namely the callable range accrual, where Darwinian model risk can be brought to light mathematically and quantified numerically. 
The consideration of the bad vs.\ not-so-bad traders allows assessing the relevance of the proposed HVA and KVA metrics in terms of their sensitivities to the specification of the setup.}

\subsection{Setup}    
\label{ss:se}

The risk-free asset is chosen as the numéraire.
We work in a probabilistic setup $(\Omega, \mathcal{A}, \bF  , \Q )$, where
$\bF = (\bF_t)_{t \in \bT}$ is a continuous-time
filtration with $\bT = [0,T]$
or a discrete-time one with $\bT = 0\,..\,T$, for a finite time horizon $T > 0$ (assumed integer in discrete time), interpreted as the final maturity of the portfolio of a bank;
the fininsurance probability measure $\Q$ is the hybrid of pricing and physical probability measures defined in \cite[Proposition 4.1]{ArtznerEiseleSchmidt22}, advocated in \citet[Remark 2.3]{CrepeyHoskinsonSaadeddine2019} for XVA computations. 
We assume the bank and its counterparty default-free, referring
the reader to \citet*[Section 5]{AlbaneseCrepeyBenezet22} for the addition of unhedgeable counterparty credit risk features to our setup. 
For an integrable semimartingale (in a \cadlag version if in continuous time, implicitly) \(\mathcal{Y} = (\mathcal{Y}_t)_{t \in [0,T]}\) starting at $0$, interpreted as cumulative cash flow process of some financial derivative, we define its fair value process \(Y = va(\mathcal{Y})\), resp.\ its fair callable %
(at constant recovery $R \in [0,1]$) value process $Z = \widetilde{va}(\cY)$ (for $\mathcal{Y}$ uniformly integrable over $\bT$), by 
\begin{align}
    \label{e:values}
    Y_t &= \mathbb{E}_t \left[ \mathcal{Y}_T - \mathcal{Y}_t \right]\sp t \in \bT,\mbox{ resp.} \\
    \label{e:amvalues}
    Z_t &=   \esssup_{\tau \in \cT^t} \mathbb{E}_t \left[ \cY_{\tau} - \cY_t + R Z_{\tau} \right]\sp t \in \bT,
\end{align}
where \(\mathbb{E}_t\) is the conditional expectation operator with respect to \(\bF_t\) under the measure \(\Q \), $\cT^t$ denotes the set of  the $[t,T] \cap \bT$ valued $\bF$ stopping times, and $R$ in $[0,1]$ is a recovery rate upon call (we will quickly reduce our attention to a standard optimal stopping setup where $R=0$). 
In particular, \(Y + \mathcal{Y}\) is a martingale and, assuming \eqref{e:amvalues} well posed (e.g. for $R=0$),
$Z+\cY$ is a supermartingale.
We denote by $\bR_t, \VaR_t$ and $\ES_t$ the $(\bF_t,\Q)$ conditional probability, value-at-risk (at some given confidence level $\alpha \in (\frac12,1)$ which is fixed throughout the paper) and expected shortfall (in the tail conditional expectation sense of an expected loss given this loss exceeds its value at risk). 

As in \cite{AlbaneseCrepeyBenezet22}, we consider a dual-model environment: on one side, a global, fair valuation model (akin to the ``reference model" advocated for model risk assessment in \citet*{barrieu2015assessing}), in which European prices (resp.\ prices of callable assets) are value processes as per \eqref{e:values} (resp.\ callable value processes as per \eqref{e:amvalues}) of the corresponding cash flow; on the other side, local models used by traders for handling their deals. Due to the use of a local model \b{(even recalibrated at all times to the fair valuation one)}, the raw profit-and-loss process of a deal (raw in the sense of not accounting for model risk reserves) is not a $\Q$ martingale in the global model, not even in the case of a European deal.

\brem \b{The use and (re)calibration of local models plays a central role throughout this work -- from the deviation of the raw profit-and-loss process from $\Q$ martingality, to the explosion of the local model, when calibration becomes impossible. For simplicity, we assume in this study that calibration is either perfect or impossible. This is satisfied in our numerical example, where analytic formulas for calibrated parameters and explicit explosion times of the local model are available. We do not analyze intermediate cases in which the trader would employ a poorly calibrated model, whether this is due to infrequent recalibration or approximate calibration via numerical optimizers. 
Considering such intermediate calibration scenarios would substantially increase theoretical and numerical complexity, whilst obscuring the financial interpretation.} \erem

These deviations from martingality due to the use of local models by traders deserve a risk-adjusted reserve, so that the  profit-and-loss process of the bank adjusted for the reserve becomes a submartingale 
in line with a remuneration of the shareholders of the bank at some hurdle rate $h$ (e.g.\ $10\%$). 
We proceed in two steps. First, an HVA reserve computable deal by deal (or at the hedging set level if it also accounts for dynamic hedging transaction costs the way addressed in \citet[Section 3]{AlbaneseCrepeyBenezet22}) makes the profit-and-loss a martingale at each deal (or hedging set) level. Second, the profit-and-loss of the bank is risk-adjusted by a KVA reserve, computed at the level of the balance-sheet of the bank as whole.

\brem
\b{A related concern is about the possibility of double-counting risk when incorporating multiple valuation adjustments.  In the present framework, the risk of double-counting among valuation adjustments is carefully mitigated by the distinct roles and scopes of each adjustment. The HVA specifically quantifies and centers the expected losses arising from model risk and suboptimal hedging strategies or exercise decisions,  adjusting the valuation distribution to reflect these realized or anticipated losses. In contrast, the KVA is designed to cover the tail risk by providing compensation for extreme loss scenarios that exceed expected losses. %
Because the HVA and KVA address different aspects of risk---expected losses versus tail losses---and operate on separate layers of the risk distribution, they are complementary rather than overlapping. Thus, the framework avoids double-counting by ensuring that the HVA corrects for average model-related losses, while the KVA addresses capital costs for extreme risk, leading to a consistent and additive valuation adjustment structure.}
\erem
   
\subsection{Outline and Notation}    
\label{ss:ou}

Section \ref{s:call} is dedicated to the definition and theoretical study of these $\HVA$ and $\KVA$ in the case of (derivatives portfolios including) callable claims. We detail two particular cases, associated to the afore-mentioned bad and a  not-so-bad trader.
We then illustrate the theory by computing the corresponding metrics for a stylized but ``typical'' structured product (callable range accrual) in a discrete time setting. Section \ref{s:dt} introduces the product and specifies associated global and local models. Section \ref{s:num}  provides a detailed numerical analysis and interpretation of the reserves, decomposing them across factors such as valuation switch, suboptimal exercise, and incorrect hedging strategy. It turns out that the risk-adjusted reserve for model risk can be significantly more substantial than the mere valuation difference between models. This highlights the importance of accounting for the misspecification of hedging and exercise strategies in model risk reserves. Section \ref{s:concl} concludes.

We write $\mathbf{1}_A$ and $\ind_A$ for the indicator of a deterministic or random set $A$, and $x^{\pm}=\pm x \mathbf{1}_{\pm x> 0}$ for any real $x$. We denote by
$\boldsymbol\delta_{\theta} $, a Dirac measure at time $\theta $, and by
$X^{\theta}$, a process $X$ stopped at time $\theta $.

\section{HVA for Callables: Abstract Framework}\label{s:call}

This section introduces our dual-model setup, similarly as in \cite{AlbaneseCrepeyBenezet22}. We refer to the later sections of the paper for a detailed example with illustrative numerics.

We assume that the bank buys a callable structured product from a client. The trader of the bank uses a local model to price and statically hedge the deal. At least, this holds up to a positive stopping time $\ts \in \cT^0$, called model switch time, at which, if the deal has not yet been terminated, the traders starts using the global model. In this scenario, new hedging ratios are computed and the static hedging strategy is rebalanced at $\ts$. As in \citet*{AlbaneseCrepeyBenezet22}, one could add to the trader's strategy a dynamic hedging component. Our focus in this work is on the static hedging side, hence we refrain from doing so, to alleviate the notation.  
 
 Denoting by $\tex \in \cT^0$ an exercise time chosen by the trader for the asset, the raw profit-and-loss process of the trader is thus given, for $t <\tex$ in $ \bT$, by  
\beql{e:PnL genprel}
\pnl_t =& \cQ_{t} + \theq_{t} \indi{t<\ts} + \theQ_{t} \indi{{t}\ge\ts} - \theq_0 \\ 
&- \left(\cP^{loc}_{t\wedge\ts} + p^{loc}_{t\wedge\ts} - p^{loc}_0\right) - \indi{t \ge \ts}\left(\cP^{fair}_{t} - \cP^{fair}_{\ts} + P^{fair}_{t} - P^{fair}_{\ts}\right),
\eeql
where:
\begin{itemize}
    \item $\cQ$ denotes the cumulative cash flow process promised by the client to the bank through the deal,
    while $\cP^{loc}$ (resp.\ $\cP^{fair}$) denotes the cumulative cash flow process promised by the bank to the hedging markets through a static hedging component constructed at time $t=0$ (resp.\ constructed at time $t=\ts$); the processes $\cQ$, $\cP^{loc}$ and $\cP^{fair}$ are assumed to be integrable (uniformly over $\bT$, regarding $\cQ$), optional, and stopped at $T$;
    \item $\theq$ (resp.\ $\thep^{loc}$) is the price of the deal (resp.\ of its time-$0$ static hedging component, assumed European), computed by the trader of the bank in the setup of a local model used for pricing and hedging the deal before the stopping time $\ts$;
    \item $\theQ = \widetilde{va}(\cQ)$ (resp.\ %
    $\theP^{fair} = va(\cP^{fair})$) is the fair callable value
    of the deal (resp.\ the fair value of its time-$\ts$ static hedging component, assumed European), used by the trader of the bank from time $\ts$ onwards.%
\end{itemize} 
The formula \eqref{e:PnL genprel} is similar to \citet*[Eqn.\ (2)]{AlbaneseCrepeyBenezet22},
\begin{itemize}
\item with $Q$, the fair callable value of $\cQ$ here, instead of its fair value there, 
\item with, for $t \in \bT$, 
the abstract quantities $\cP_t$, $p_t$ and $P_t$ 
there specified as
$\cP_t := \cP^{loc}_{t \wedge \ts} + \indi{t \ge \ts} (p^{loc}_{\ts}-P^{fair}_{\ts} + \cP^{fair}_t - \cP^{fair}_{\ts})$, $p_t \indi{t<\ts} = p^{loc}_t \indi{t<\ts}$, and $P_t \indi{t \ge \ts} :=  P^{fair}_t \indi{t\ge\ts}$ here, 
\item and here without the dynamic hedging component there. 
\end{itemize}
Moreover, in the present setup of a callable asset with recovery rate $R \in [0,1]$, the raw profit-and-loss process may additionally jump at the exercise time $\tex$, by the amount (cf.\ \eqref{e:PnL genprel})
  \beql{e:pnlj}
  &\pnl_{\tex}- \pnl_{\tex-}=\cQ_{\tex}-\cQ_{\tex-} +R \left( q_{\tex}\indi{\tex<\ts} +Q_{\tex}\indi{\tex\ge\ts}\right)\\&\qqq- q_{\tex-}\indi{\tex\le\ts} - Q_{\tex-}\indi{\tex>\ts} \\&\quad
  - \left(\cP^{loc}_{\tex}\indi{\tex<\ts} + \cP^{fair}_{\tex}\indi{\tex\ge\ts} - \cP^{loc}_{\tex-}\indi{\tex\le\ts} - \cP^{fair}_{\tex-}\indi{\tex>\ts}\right) \\&\quad
  - \left( p^{loc}_{\tex}\indi{\tex<\ts} + P^{fair}_{\tex}\indi{\tex\ge\ts} - p^{loc}_{\tex-}\indi{\tex\le\ts} - P^{fair}_{\tex-}\indi{\tex>\ts} \right) ,
\eeql
where $\thep^{loc}_{\tex}\indi{\tex< \ts} + \theP^{fair}_{\tex}\indi{\tex\ge \ts}$  is the liquidation cash flow of the hedge, assumed liquidly tradable at all times.
A recovery rate $R=1$ on the asset upon call would mean that the asset is liquidly sold at $\tex$; $R<1$ covers the more realistic case of a structured product that is illiquid and can only be called by the bank for a fraction of its value at $\tex$.
 
Hereafter we assume $R=0$, i.e.\ the asset is callable at zero recovery. In particular, from \eqref{e:amvalues} with $R=0$, the classical theory of optimal stopping (see e.g. the seminal works \citet{neveu1975discrete}, for the  discrete time case, and \citet[Chapter II]{ElKaroui1981SaintFlour}, in continuous time) indicates that, at any time $t \in \bT$, an optimal exercise time starting from $t$ for the problem $\theQ = \widetilde{va}(\cQ)$ is given by 
\beql{e:tau t}
\tau^t := \inf \left\{ \bT \ni s \ge t ; Q_s=0\right\} \wedge T.
\eeql
In addition, $\cQ+\theQ$ is a supermartingale, and we denote by $K$ its drift, i.e.\ the unique nondecreasing integrable predictable process such that $K_0=0$ and $\cQ+\theQ+K$ is a martingale.

Gathering \eqref{e:PnL genprel} and \eqref{e:pnlj} for $R=0$, we obtain the following:
\bd\label{def}
The raw profit-and-loss process of the trader is given, for all $t \in \bT$, by
\beql{e:PnL gen}
    \pnl_t =& \cQ_{t\wedge\tex} + \theq_{t\wedge\tex} \indi{t\wedge\tex<\ts} + \theQ_{t\wedge\tex} \indi{{t\wedge\tex}\ge\ts} - \theq_0 \\ 
    &- \left(\cP^{loc}_{t\wedge\tex\wedge\ts} + p^{loc}_{t\wedge\tex\wedge\ts} - p^{loc}_0\right) - \left(\cP^{fair}_{t\wedge\tex} - \cP^{fair}_{\ts} + P^{fair}_{t\wedge\tex} - P^{fair}_{\ts}\right)\indi{t\wedge\tex \ge \ts}
    \\ &- \indi{  t \ge \tex} \left(\indi{\tex<\ts}\theq_\tex+\indi{\tex\ge\ts}\theQ_\tex\right).
\eeql 
\ed

\brem\label{rem:non mgale}
{\rm\emph{(i)}}{\def\ba{*} Because of model risk, $\pal$ fails to be a 
martingale, as opposed to the model-risk-free version of  \eqref{e:PnL gen},
\bel%
\cQ_{t\wedge\tau_e^{\ba}}   + \theQ_{t\wedge\tau_e^{\ba}}  - \theQ_0  - \left(\cP^{\ba}_{t\wedge\tau_e^{\ba}}   + \theP^{\ba}_{t\wedge\tau_e^{\ba}}   - \theP^{\ba}_0\right)%
\eel
that would result from using only the global model everywhere (assuming optimal exercise $\tau_e^\star = \tau^0$ as per \eqref{e:tau t} such that, in particular $\theQ_{\tau_e^{\ba}}=0$).
In the equation above,
we use $\ba$ to emphasize that the hedges and exercises decisions computed within the global model would differ from the ones in \eqref{e:PnL gen}, which are computed within the trader's local model at time $t=0$. }  

\noindent{\rm \emph{(ii)}}  There may also be American claims puttable by the clients, as opposed to callable by the bank (with zero recovery for notational simplicity) in the paper. In the case of puttable claims, we conservatively assume that they are optimally exercised by the clients, without benefit for the bank, so that we do not need to introduce the corresponding ``nonincreasing processes" that would play a role symmetrical to our nondecreasing processes $K$ for callable claims.
\erem

We make the following natural assumption regarding the local model used by the trader.
\bhyp\label{hyp:recalib}
For all $t \in \bT$, on $\{t<\ts\}$, the local model is calibrated to the static hedging instruments' fair prices, i.e.\ one has 
\beql{e:pq}
p^{loc}_t = P^{loc}_t := va(\cP^{loc})_t.%
\eeql
In addition, we assume that $q_t=q^t_t$,
for some price process $(q^t_s)_{t \le s \in \bT}$ of the asset computed in a local model calibrated at time $t$ to the fair valuation of the hedging assets (assumed European).\\
Last, we assume that, for each $t \in \bT$, on $\{t<\ts\}$, in the time-$t$ calibrated local model, an optimal stopping time for the callable deal
is given by
\beql{eq:theta t}
\theta^t := \inf \left\{ s \in [t,T) \cap \bT ; q^t_s=0\right\} \wedge T.
\eeql
\ehyp
\noindent
Note that $q_t$ and $p^{loc}_t$ are completely unspecified on $\{t \ge \ts\}$, but in view of \eqref{e:PnL gen} they are irrelevant on this set.

\citet*{AlbaneseCrepeyBenezet22} was restricted to European-style structured products with $\tex$ constrained to be identically $T$ in \eqref{e:PnL gen}, considering various static and/or dynamic hedging strategies in this setup. In the present paper, instead, we  play with various stopping times $\tex$ reflecting  optimal calls by the trader from the viewpoint of different models, also depending on the trader's ability and willingness to switch to the fair valuation model if his local model no longer calibrates.

\subsection{Hedging Valuation Adjustment}
 The HVA is a defined as a reserve imposed by the bank to the trader to cope with misvaluation model risk, so that the $\HVA$-compensated pnl, $\pal-\HVA + \HVA_0$, is a
martingale (see Remark \ref{rem:non mgale}(i)):

\bd The hedging valuation adjustment ($\HVA$) is 
\beql{e:HVA def}
\HVA \eqdef -va( \pal).
\eeql
\ed
    
\bp\label{p:HVA}
Under Assumption \ref{hyp:recalib}, we have, for all $t \in \bT$, 
\beql{e:HVA gen}
&\HVA_t  = 
\left(\theq_{t\wedge\tex} -\theQ_{t\wedge\tex}\right) 
\indi{t\wedge\tex<\ts} 
- \espc{t}{ 
\left(\theq_{\tex}-\theQ_\tex\right) 
\indi{\tex<\ts}} \\ &  \qqq +  \left(\cQ+\theQ\right)_{t\wedge\tex} - \espc{t}{\cQ_\tex+Q_\tex} \\ &  \qqq + \indi{t<\tex}\espc{t}{\indi{\tex<\ts}\theq_{\tex}+\indi{\tex\ge\ts}Q_{\tex}}. 
\eeql
\ep
\proof
Recall $X^{\tex}:=X_{\cdot\wedge \tex}$. Let $t\in\bT$. Under Assumption \ref{hyp:recalib}, each of the two parentheses in the second line in \eqref{e:PnL gen}, hence this second line as a whole, is a zero-valued martingale. Therefore
\bel
&va(pnl)_t = va\left(\left(\cQ+\theq\indi{\cdot<\ts}+\theQ\indi{\cdot\ge\ts}\right)^{\tex}\right)_t%
- va\left(\indi{\cdot\ge\tex}\left(\indi{\tex<\ts}\theq_\tex+\indi{\tex\ge\ts}\theQ_\tex\right)\right)_t \\
=& va\left(\left(\cQ+\theQ+K+\left(\theq-\theQ\right)\indi{\cdot<\ts}-K\right)^{\tex}\right)_t %
- va\left(\indi{\cdot\ge\tex}\left(\indi{\tex<\ts}\theq_\tex+\indi{\tex\ge\ts}\theQ_\tex\right)\right)_t \\
=& va\left(\left(\left(%
\theq-\theQ\right)%
\indi{\cdot<\ts}\right)^{\tex}\right)_t - va(K^{\tex})_t %
- va\left(\indi{\cdot\ge\tex}\left(\indi{\tex<\ts}\theq_\tex+\indi{\tex\ge\ts}\theQ_\tex\right)\right)_t,
\eel
where we also used that $\cQ+\theQ+K$ (with $K$ as introduced in the third bullet point after \eqref{e:PnL genprel}) is a martingale. %
Thus, by \eqref{e:HVA def}, 
\beql{e:hva parlant}
\HVA_t =& va\left(\left(\left(%
\theQ-\theq\right)%
\indi{\cdot<\ts}\right)^{\tex}\right)_t %
+ va(K^{\tex})_t + va\left(\indi{\cdot\ge\tex}\left(\indi{\cdot<\ts}\theq+\indi{\cdot\ge\ts}\theQ\right)\right)_t.
\eeql
We now compute each term separately in \eqref{e:hva parlant}. 
First we have by \eqref{e:values}, since also $\tex \le T$,
\bel
va\left(\left(\left(%
\theQ-\theq\right)%
\indi{\cdot<\ts}\right)^{\tex}\right)_t &= \espc{t}{\left(\left(%
\theQ-\theq\right)%
\indi{\cdot<\ts}\right)_T^{\tex}} - \left(\left(%
\theQ-\theq\right)%
\indi{\cdot<\ts}\right)_t^{\tex} \\
&= %
\left(\theq_{t\wedge \tex}-\theQ_{t\wedge \tex}\right)\indi{t\wedge \tex<\ts} - \espc{t}{\left(\theq_{\tex}-\theQ_{\tex}\right)\indi{\tex<\ts}}
\eel
Next, since $\cQ+\theQ+K$ is a martingale and $T\wedge\tex=\tex$, we have
\bel
va(K^{\tex})_t = -va((\cQ+\theQ)^{\tex})_t = (\cQ+\theQ)^{\tex}_t - \espc{t}{\cQ_\tex+\theQ_\tex}.
\eel
Last, since $T\wedge\tex=\tex$, we compute
\bel
&va\left(\indi{\cdot\ge\tex}\left(\indi{\tex<\ts}\theq_\tex+\indi{\tex\ge\ts}\theQ_\tex\right)\right)_t \\ &= \espc{t}{\indi{T\ge\tex}\left(\indi{\tex<\ts}\theq_\tex+\indi{\tex\ge\ts}\theQ_\tex\right)}-\indi{t\ge\tex}\left(\indi{\tex<\ts}\theq_\tex+\indi{\tex\ge\ts}\theQ_\tex\right) \\
&= \espc{t}{\indi{\tex<\ts}\theq_\tex+\indi{\tex\ge\ts}\theQ_\tex}-\indi{t\ge\tex}\left(\indi{\tex<\ts}\theq_\tex+\indi{\tex\ge\ts}\theQ_\tex\right) \\
&= \espc{t}{\indi{t<\tex}\left(\indi{\tex<\ts}\theq_\tex+\indi{\tex\ge\ts}\theQ_\tex\right)}.\ \finproof
\eel

\brem\label{rem: interp}
At time $t=0$, the bank pays $\theq_0$ to its client. In addition, 
through the first term of the HVA \eqref{e:HVA gen} valued at $t=0$, the client pays $q_0-Q_0$. At this stage, from the viewpoint of the bank it is as if the bank had paid $\theQ_0$ to the client, i.e.\ the fair valuation price is restored.
So the first HVA term in \eqref{e:HVA gen} is a reserve compensating the misvaluation before the model switch. The other terms
are reserves for potentially suboptimal exercise.
\erem
\subsection{Capital Valuation Adjustment}
While the fair valuation prices are restored via the HVA (see Remark \ref{rem: interp}), the hedge is still computed in the local model before the model switch, hence it can only be wrong and leave some (or even enhance)  market risk, which is not taken into account through the HVA. Similarly, a reserve for 
suboptimal exercise is provided, but the corresponding risk is not hedged. Unhedged risk requires shareholder's capital to cover the losses $-\pal+\HVA-\HVA_0$ associated with the still wrong hedge and exercise policy. The level of capital at risk of the bank is assumed to target a certain economic capital. The bank then needs to remunerate shareholders at some hurdle rate on their capital at risk. Under a cost-of-capital approach to the management of financial derivatives, the reserve for model risk therefore needs to be risk-adjusted, in the form of a related contribution to the capital valuation adjustment (KVA) of the bank, which is the amount needed by the bank for remunerating its shareholders for their risk.

We now define the corresponding economic capital ($\EC$) and the associated capital valuation adjustment ($\KVA$) processes of the bank.  
    \begin{defi}\label{d:ckvacont} 
    For all $t \in \bT$, we set\footnote{cf.~\citet*[Section 4.1]{AlbaneseCrepeyBenezet22}.}
    \beql{e:eckvacont}
    & \EC_t=\ES_{t}\big[ -(\pal_{(t+1)\wedge\Tb}-\pal_t)+\HVA_{(t+1) \wedge \Tb} -\HVA_t \big],\\
    & \KVA_t=\hurdle \E_t\left[\int_{t}^{\Tb} e^{-\hurdle (s-t)}  \max(\KVA_s,\EC_s)  \boldsymbol\mu_t(ds)\right],
    \eeql 
    for some positive and constant hurdle rate $\hurdle$ (set to 10\% in our numerics), and where $\boldsymbol\mu_t$ is the Lebesgue measure on $[t,T]$, if $\bT=[0,T]$, or $\boldsymbol\mu_t = \sum_{s=t+1}^T \boldsymbol\delta_{s}$ , if $\bT =  0\,..\,T$. 
     \ed 
\noindent
This specification ensures 
that the bank has exactly enough KVA\footnote{at least in the continuous time setup where $\bT=[0,T]$, cf.\ \citet*[Remark 2.6]{Crepey21}.}
to remunerate its shareholders at the target hurdle rate $\hurdle$ on their capital at risk, dynamically in time.
     \brem In this work, in order to focus on the model risk associated to the use of local models and their impact on hedges and exercise strategies, we assume that the bank's portfolio is reduced to one product and its hedge. In general
     the economic capital and the KVA can only be computed at the level of the bank's portfolio.
     \erem

\subsection{The Bad and Not-So-Bad Traders}

In what follows we specify the above to the special cases of the bad and the not-so-bad trader introduced in Subsection \ref{ss:da}. The two traders behave similarly from a hedging perspective, but they differ in their early exercise strategies. Hereafter, we denote by $\texbad$ and $pnl^{bad}$ (resp.\ $\texnsb$ and $pnl^{nsb}$) the exercise time and the raw pnl of the bad (resp.\ not-so-bad) trader.

We assume that before $\ts$ the bad trader aims at exercising optimally with respect to the local model by considering the stopping time
    \beql{eq:theta star}
    \theta^\star &:= \inf \left\{ t \in [0, \ts)\cap\bT;\, \theta^t=t\right\} \wedge \ts
     =
     \inf \left\{ t \in[0, \tau_s )\cap\bT;\, \theq_t = 0\right\}\wedge\tau_s ,
    \eeql
where $\theta^t$ is the optimal exercise time of the trader computed in the time-$t$ calibrated model as per Assumption \ref{hyp:recalib}, and where the equality holds by definition of $q$ after \eqref{e:pq}.
But if the local model no longer calibrates before the 
asset reaches zero value in the local model, i.e.\ if $\theta^\star=\tau_s$, then
the bad trader is unable or unwilling to reshuffle his hedge according to the prescriptions of the global model; if his position is still open at $\ts$, he just closes it at that time by calling the asset and unwinding the hedge under the pricing terms of the fair valuation model.  Accordingly:
\bd\label{ex: bad trader} The exercise policy of the bad trader is given by 
    \beql{eq:optimal stopping}
        & \texbad:= \theta^\star\le \ts. 
    \eeql
\ed
\noindent Since $\texbad \le \ts$, we obtain from 
\eqref{e:PnL gen} that
\beql{e:bad pnl}
\pnl^{bad}_t =& \cQ_{t\wedge\texbad} + \theq_{t\wedge\texbad} \indi{t\wedge\texbad<\ts} + \theQ_{t\wedge\texbad} \indi{{t\wedge\texbad}=\ts} - \theq_0 \\ 
&- \left(\cP^{loc}_{t\wedge\texbad} + P^{loc}_{t\wedge\texbad} - P^{loc}_0 \right) 
- \indi{  t \ge \texbad} \left(\indi{\texbad<\ts}\theq_{\texbad}+\indi{\texbad=\ts}\theQ_{\texbad}\right). 
\eeql

The not-so-bad trader behaves as the bad trader before the explosion time $\ts$ of the local model. However, if $\ts$ occurs before the termination of the deal, then the not-so-bad trader switches to the global model at $\ts$, after which he aims at exercising optimally according to the latter,  considering the stopping time
\beql{e:thetaust}
   \tau^\star \eqdef \tau^{\ts}
, 
\eeql
where $\tau^{\ts}$ is the optimal exercise time of the trader computed in the global model at time $\ts$ as per \eqref{e:tau t}. 
As such:
\bd\label{ex: nsb trader}  The exercise time of the not-so-bad trader is given by
\beql{e:thetau}
  \texnsb := {\theta^{\star} \indi{\theta^{\star}<\ts} + \tau^{\star} \indi{\theta^\star = \ts}},
\eeql 
with $\theta^{\star}$ as per \eqref{eq:theta star}-\eqref{eq:theta t} and $\tau^\star$ as per
\eqref{e:thetaust}.
The raw pnl of the not-so-bad trader is then given by \eqref{e:PnL gen} with $\tex=\texnsb$, for which both $\cP^{loc}$ and $\cP^{fair}$ are material in \eqref{e:PnL gen} (in contrast to the bad trader for which $\cP^{fair}$ is irrelevant, see \eqref{e:bad pnl}).
\ed

\section{Stylized Callable Range Accrual in Discrete Time\label{s:dt}}

In the sequel of the paper, we consider a stylized
callable range accrual in discrete time $\bT = 0\,..\,T$ with $T$ positive integer, in the natural augmented filtration $\F=\F^N$ of a process $N = (N^{}_l)_{0\le l\le\Td  }$  such that $N_0=0$ and $N_{l+1} - N_{l}$ is an independent Poisson random variable with parameter $\h\gamma^{}_{l}\ge   0$, for each $l \in 0\,..\,\Td -1$.
The range accrual cumulative cash flow process is defined by
\beql{e:rangeaccr disc}
\cQ^{}_{k} \eqdef \h \sum_{l=1}^k\left(\indi{I^{}_l = -1} - \indi{I^{}_l = 1} \right), \quad k \le\Td ,
\eeql
where
\beql{e:I}
& I^{}_k = I^{}_0 (-1)^{N^{}_k} = I^{}_l (-1)^{N^{}_k - N^{}_l}\sp 0\le  l \le k \le\Td.
\eeql
This process $I$ plays the role of the global model in our example.

At each time $0 \le k \le T$,  the role of the local model is then played by the process $i^{k} = (i^k_l)_{l=k}^T$ such that, for $k \le l \le T$, 
\bel
i^k_l = \begin{cases}
1 \text{ if } i^k_k = 1 \text{ and } n^k_l = 0, \\
-1 \text{ otherwise, i.e.\ if } i^k_k = -1 \text{ or } n^k_l \ge 1,
\end{cases}
\eel
where $n^{ k} = (n^{ k}_l)_{k \le l \le\Td }$ is a process with independent increments such that $n^{ k}_k=0$ and $n^{ k}_{l+1} - n^{ k}_{l}$ is an independent Poisson random variable with some parameter $\nu^{ k}_{l}$, for each $l\in k\,..\, T-1$. 
The parameters
$\nu^{ k}_{l} , l\in k  \,..\,\Td -1$,
are recalibrated at each time $k$ (as long as it is possible) to the time-$k$ fair values $P^{\mkt}_k (\ell)$ of the binary options with payoff $\indi{I_\ell=-1}$, $   \ell \in  k\,..\, T$, which will be used a static hedging assets for the claim (see Assumption \ref{e:hypcont} below).
Note that $P^{\mkt}_k(k) = 0$ (resp.~$P^{\mkt}_k(k) = 1$) if $I^{}_k=+1$ (resp.~$I^{}_k=-1$).

\brem
{\rm
\emph{(i)}} In the market a typical range accrual pays a reference rate to the bank whenever this rate is outside a corridor. Our event $\{I_t=1\}$ mimics the normal situation where the underlying rate would be inside the corridor at time $t$, while the event $\{I_t=-1\}$ corresponds to the extreme case where the rate would be outside the corridor. We use this simple example as a proxy to investigate the features of model risk that may have been responsible for huge losses in the structured product crises mentioned in the introduction of the paper. Namely, the bank which buys the product is long the extreme event on the asset side, but also accounting for its misspecified hedge, it will end-up short the extreme event. 
This is the key picture  
that we want to capture in our setup.
 {\rm
\hfill\break
\emph{(ii)}} In the local model, whenever the extreme event (which the bank is long of on the asset side) occurs, then it persists until maturity. Hence the local model puts more weight on the scenarios that benefit to the bank on the asset side. In particular, the premium of the asset computed in the local model will be higher than the one in the fair valuation model (cf.\ Figure \ref{fig:fct-q-Q} in Section \ref{s:num}). This induces an attractive price for the client selling the asset to the bank, which is the source of ``Darwinian model risk'' (of adverse model selection) in \cite{AlbaneseCrepeyIabichino19}: see Section \ref{ss:da}.
 {\rm
\hfill\break
\emph{(iii)}} As we are in discrete time and that the processes $I$ and $i^k$ can only take two values $\pm 1$, our setup is amenable to exact numerics, without Monte-Carlo simulation or PDE approximation biases (see Sections \ref{s:dt} and \ref{s:num}).
\erem

Hereafter in this section, we study the theoretical properties of our stylized range accrual.

    \subsection{Pricing and Hedging}\label{s:price hedge}

  \subsubsection{Hedging assets and calibration of the local model}
    
We have the following two lemmas regarding the pricing of the binary options in the fair valuation and local models. These binary options being used as calibration and static hedging assets, we deduce as a corollary the calibration of the local model to these fair valuation prices.

    \bl
    The time-$k$ fair valuation price of the binary option with maturity $\tell$ is given, for each $0 \le k \le \ell \le\Td $, by
    \beql{l:epT disc}
    P_k(\ell)= \indi{I^{}_k =1} \frac{1 - e^{-2\h \sum_{l=k}^{\ell-1} \gamma^{}_l }}{2}+\indi{I^{}_k =-1} \frac{1 + e^{-2\h \sum_{l=k}^{\ell-1} \gamma^{}_l}}{2} .
    \eeql
    \el
    \proof
    We compute
    \bel
    P_k(\ell) &= \espc{k}{\indi{I^{}_\ell=-1}} = \espc{k}{\indi{I^{}_k (-1)^{N^{}_\ell -N^{}_k} = -1}} \\
    &= \indi{I^{}_k=1}\espc{k}{\indi{(-1)^{N^{}_\ell-N^{}_k}=-1}} + \indi{I^{}_k=-1}\espc{k}{\indi{(-1)^{N^{}_\ell -N^{}_k+1}=-1}} \\
    &= \indi{I^{}_k=1} \prob{N^{}_\ell - N^{}_k \mbox{ odd}} + \indi{I^{}_k=-1} \prob{N^{}_\ell - N^{}_k \mbox{ even}}, 
    \eel which yields \eqref{l:epT disc}.~\finproof
    
    \def\kappa{l}
    
    \bl For $ 0 \le k \le \ell \le\Td $,  the time-$k$ local model price of the binary option with maturity $\tell$ is 
    \bel
    \espc{k}{\indi{i^{ k}_\ell = -1}} &= \indi{i^{ k}_k=-1} + \indi{i^{ k}_k=1}(1 - e^{- \sum_{\kappa=k}^{\ell-1} \h \nu^{ k}_\kappa}).
    \eel
    \proof
    We compute
    \bel
    \espc{k}{\indi{i^{ k}_\ell = -1}} &= \indi{i^{ k}_k=-1} + \indi{i^{ k}_k=1}\espc{k}{\indi{n^{ k}_\ell \ge 1}} \\
    &= \indi{i^{ k}_k=-1} + \indi{i^{ k}_k=1}\left(1 - \prob{n^{ k}_\ell = 0}\right),  
    \eel
    where $\prob{n^{ k}_\ell = 0}= e^{- \sum_{\kappa=k}^{\ell-1} \h \nu^{ k}_\kappa}$.
    \finproof
    \el 
    \bcor\label{cor:calib} Assuming $I_0=1$, as long as $I_k=1$, the local model calibrates to the term structure $P_k(\cdot)$ in \eqref{l:epT disc} via $i^k_k = I_k=1$ and 
    \beql{eq:calib disc}
    1 - e^{-\sum_{\kappa=k}^{\ell-1} \h \nu^{ k}_\kappa} = P_k(\ell)\sp k<\ell
     \mbox{ , i.e.~} \h \nu^{ k}_{\ell-1} = - \ln(1 - P_k(\ell)) - \sum_{\kappa=k}^{\ell-2} \h \nu^{ k}_\kappa\sp k<\ell 
     .
    \eeql
    As soon as the extreme event occurs, i.e.\ at 
    \beql{e:thetausdis}\ts = \inf \left\{k \in 0\,..\,T;\, I_k = -1\right\}\wedge T,\eeql 
    the trader's local model no longer calibrates (at least if $\ts<T$, and note that we neither need nor use any model at time $T$).
    \ecor 

   \subsubsection{Range accrual and its hedging ratios}
   
    We now compute, for each $0 \le t \le T$, the fair valuation and local prices of the  range accrual. Let, for $1 \le l \le T$,
    \beql{e:uv}&\even_l= \prob{N^{}_l - N^{}_{l-1} \mbox{ even}} =\frac12(1 +e^{-2 \h\gamma^{}_{l -1}}),\\&\odd_l =\prob{N^{}_l - N^{}_{l-1} \mbox{ odd}} =\frac12(1 - e^{-2 \h\gamma^{}_{l-1}}).\eeql

    \bp\label{prop: pricing hedging fair model}
    The fair callable value of the range accrual at time $k \le\Td $ is equal to
    \begin{align}
    \label{eq:vaQ disc} Q^{}_k & =\h \sum_{\ell=k+1}^{\Td} \Big(\Athe^{}_k(\ell) P_k(\ell) - \theB^{}_k(\ell) \big(1-P_k(\ell)\big)\Big),
    \end{align}
    with
    \beql{e:AthetheB}
    \Athe^{}_k(\ell) \eqdef \frac{\espc{k}{\indi{I^{}_\ell=-1}\indi{ \tell \le \tau^{ k}}}}{P_k(\ell)} \sp \theB^{}_k(\ell) \eqdef  \frac{\espc{k}{\indi{I^{}_\ell=1}\indi{ \tell \le \tau^{ k}}}}{1-P_k(\ell)}, k 
    \le \ell \le T,
    \eeql
    where $\tau^{ k}$ is the optimal exercise time computed at time $k$ as per \eqref{e:tau t}.
    
   The process $Q$  in \eqref{eq:vaQ disc} can be represented as $Q^{}_k = Q^{}(k,I^{}_k)$, for the pricing function $Q^{} : \{0,\dots,\Td\} \times \{1,-1\} \to \mathbb{R} $ such that
    \beql{e:Qd}
    & Q^{}(\Td,\mp 1)  = 0\mbox{  and, for } 0 \le k <\Td  ,\\
    &Q^{}(k,-1)  
    = \h e^{-2 \h\gamma^{}_k} +v_{k+1} Q^{}(k+1,1) +u_{k+1} Q^{}(k+1,-1) > 0
    , \\
    &Q^{}(k,1)    = \max\left(0, - \h e^{-2\h\gamma^{}_k}  + u_{k+1}  Q^{}(k+1,1) +v_{k+1}  Q^{}(k+1,-1)\right) . 
    \eeql
    \ep
    \proof
    We compute, with $\tau^k$ and $\Athe^{}_k(\ell)$, $\theB^{}_k(\ell)$ as introduced,
    \begin{align}\label{e:maxsup}
    Q^{}_k & = \h \esssup_{\tau \in \cT^{ k}} \espc{k}{\sum_{\ell=k+1}^{\Td} \left(\indi{I^{}_\ell=-1} - \indi{I^{}_\ell=1}\right) \indi{t_\ell \le \tau}} \\\nonumber
    & = \h\sum_{\ell=k+1}^{\Td} \espc{k}{\left(\indi{I^{}_\ell=-1} - \indi{I^{}_\ell=1}\right)\indi{t_\ell \le \tau^{ k}}} \\\nonumber
    & = \h \sum_{\ell=k+1}^{\Td} \Big(\Athe^{}_k(\ell) P_k(\ell) - \theB^{}_k(\ell) \big(1-P_k(\ell)\big)\Big), \quad  
    \end{align}
    which proves \eqref{eq:vaQ disc}.
    
Moreover, by the Markov property of $I$, the process $Q^{}$ can be represented as $Q^{}_k = Q^{}(k,I^{}_k)$, where the function $Q(\cdot,\cdot)$ satisfies the backward dynamic programming equations $Q^{}(\Td,I^{}_\Td)  = 0$ and, for $0 \le k < \Td $,
    \bel 
    & Q^{}(k, I^{}_k) = \\&\max\left(0, \probc{k}{I^{}_{k+1}=1}\left(-1%
    + Q^{}(k+1, 1)\right) + \probc{k}{I^{}_{k+1}=-1}\left(1%
    + Q^{}(k+1,-1)\right)\right),
    \eel
    i.e.~  $Q^{}(\Td,\mp 1)  = 0\mbox{  and, for } 0 \le k <\Td  ,$
    \bel  
    &Q^{}(k,-1)  =  \max\left(0,v_{k+1} \left(-1%
    +Q^{}(k+1,1)\right) +u_{k+1}  \left(1%
    +Q^{}(k+1,-1)\right)\right),\\
    &Q^{}(k,1)   =  \max\left(0, u_{k+1}   \left(-1%
    +Q^{}(k+1,1)\right) +v_{k+1}  \left(1%
    +Q^{}(k+1,-1)\right)\right)  .~\finproof
    \eel

    We have the following similar statement regarding the pricing of the claim in the time-$k$ calibrated local model, recall Assumption \ref{hyp:recalib}. The proof is similar and thus omitted.
    \bp\label{prop: pricing hedging local model}
    For each $0 \le k \le T$, the callable price of the range accrual in the local model is equal to
    \beql{e:ql}
    q^{}_k &= \h \sum_{\ell=k+1}^{\Td} \Big(a^{}_k(\ell) P_k(\ell) - b^{}_k(\ell)\big(1-P_k(\ell)\big)\Big),
    \eeql
    with
    \beql{e:ab}
    a^{}_k(\ell) \eqdef \frac{\espc{k}{\indi{i^{ k}_\ell=-1}\indi{ \ell \le \theta^{ k}}}}{P_k(\ell)} \mbox{ and } b^{}_k(\ell) \eqdef \frac{\espc{k}{\indi{i^{ k}_\ell=1}\indi{ \ell \le \theta^{ k}}}}{1-P_k(\ell)}\sp k
    \le \ell \le\Td ,
    \eeql
    where $\theta^k = \inf \left\{ \ell \ge k;\, q^k_\ell = 0\right\}$, see \eqref{eq:theta t}, is an optimal stopping rule in the time-$k$ calibrated local model.
    
The process $q$ in  \eqref{e:ql} can be represented as $q^{}_k = q^{ k}(k,i^{ k}_k) = q^k(k,I_k)$, for the pricing functions $q^{ k} : \{k,\dots,\Td\} \times \{1,-1\} \to \bR$ defined, for each $0\le k\le T$, by
    \beql{e:qd}
   & q^{ k}(\Td,\mp 1)=0\mbox{ and, for } k \le l <\Td ,\\
  &  q^{ k}(l,-1)  = \h
    \Td -l , \\
  &  q^{ k}(l,1) = \max\left(0, e^{-\h \nu^{ k}_l}\left(-1%
    +q^{ k}(l+1,1)\right) + \left(1-e^{-\h \nu^{ k}_l}\right)\left(1%
    +q^{ k}(l+1,-1)\right)\right) .
    \eeql
    \ep

In view of \eqref{eq:vaQ disc}-\eqref{e:AthetheB}, at any time $k$, a natural static hedging strategy from the global model perspective, 
dubbed fair hedge below, is to sell (resp.\ buy), for each $k<\ell\le T$, an amount $\Athe^{}_k(\ell)$  (resp.\ $\theB^{}_k(\ell)$) of binary options with payoff $\indi{I_\ell=-1}$ (resp.\ $\indi{I_\ell=1}$). 
This would in fact statically replicate the range accrual if it was not for its callability (the noncallable version of the range accrual is nothing but the collection of the binaries).
 
Likewise, in view of \eqref{e:ql}-\eqref{e:ab}, at time $k$, a natural static hedging strategy from the local model perspective, dubbed local hedge below, is to sell (resp.\ buy), for each $k<\ell\le T$, an amount $\thea^{}_k(\ell)$  (resp.\ $\theb^{}_k(\ell)$) of binary options with payoff $\indi{I_\ell=-1}$ (resp.\ $\indi{I_\ell=1}$).

Accordingly:

\bhyp\label{e:hypcont}  {\rm \textbf{(i)}} 
At time $k=0$, both traders implement the local static hedge 
 \beql{l:epT discbisprel}
 \cP^{loc}_k= \h\sum_{\ell=1}^{k} \left( \thea_0(\ell) \indi{I_\ell=-1} {- \theb_0(\ell) \indi{I_\ell=1}} \right), \quad k \ge 0.\eeql
{\rm \textbf{(ii)}} At the model switch time $k=\ts$, the bad trader unwinds its position (under the conditions prescribed by the global model), while  (if $\ts<\texnsb$) the
not-so-bad trader switches to the fair static hedge
such that 
\beql{e:cPfair}
\cP^{fair}_k= \sum_{\ell= \ts+1}^{k }\left( \Athe^{}_{\ts}(\ell)\indi{I^{}_\ell=-1} - \theB^{}_{\ts}(\ell)\indi{I^{}_\ell=1}\right) , \quad k \ge 0.
\eeql
\ehyp

\brem 
 $\cP^{loc}$ is fairly valued, for $k \ge 0$, as 
 \beql{e:Plocex}
 &P^{loc}_k = \sum_{\ell=k+1}^T \left(a_0(\ell)P_k(\ell) - b_0(\ell)(1-P_k(\ell))\right),\eeql 
 with $P_k(\ell)$ as in \eqref{l:epT disc};
in particular, \eqref{e:ql} and \eqref{e:Plocex} yield that $P^{loc}_0=q_0$, which also reads $p^{loc}_0=q_0$, by Assumption \ref{hyp:recalib}.

 $\cP^{fair}$ is fairly valued, for $k \ge 0$ and on $\{k\ge\ts\}$, as 
 \beql{e:Pfairex} P^{fair}_k = \sum_{\ell=k+1}^T \left( A_{\ts}(\ell) P_k(\ell) - B_{\ts}(\ell)(1-P_k(\ell))\right) .\eeql \erem 

The following lemma allows computing the static hedging ratios $a_0(\ell)$ and $b_0(\ell)$ for all $0< \ell\le T$.

\def\ellmax{\theta^0_{\min}}\def\ellmax{\underline{\theta}^0}
\bl
Let $\ellmax := \inf \left\{0 \le l \le\Td ;\, q^0(l,1) = 0 \right\} \wedge \Td$.
{\rm\hfill\break \textbf{(i)}} We have $\ellmax\le \theta^0  = \inf \left\{ l \ge 0;\, q^0_l = 0 \right\}$ (see Proposition \ref{prop: pricing hedging local model} and \eqref{eq:theta t}).
{\rm\hfill\break \textbf{(ii)}} For all $0 \le \ell \le \ellmax$, one has $a_0(\ell)=b_0(\ell)=1$.
{\rm\hfill\break \textbf{(iii)}}  For all $\ellmax < \ell \le T$, one has $b_0(\ell)=0$.   {\rm\hfill\break \textbf{(iv)}} For all $\ellmax < \ell \le T$, one has $a_0(\ell)=\frac{P_0(\ell_{\max})}{P_0(\ell)}$.
\el
\proof { \textbf{(i)}} Notice that $q^0(\ell,1) \neq 0$ for $\ell<\ellmax$, by definition of $\ellmax$, and $q^0(\ell,-1) = T-\ell > 0$ as $\ell < T$, see Proposition \ref{prop: pricing hedging local model}. Hence $q^0_\ell\neq0$ for all $0 \le \ell < \ellmax$, implying that $\theta^0 \ge \ellmax$.
{\rm\hfill\break \textbf{(ii)}} For $0 \le \ell \le \ellmax $, (i) implies $\ell \le \ellmax \le \theta^0$, hence \eqref{e:ab} yields
\bel
a_0(\ell) = \frac{\esp{\indi{i^0_\ell=-1}}}{P_0(\ell)} \text{ and } b_0(\ell) = \frac{\esp{\indi{i^0_\ell=1}}}{1-P_0(\ell)},
\eel
where both quantities are equal to $1$ as, by assumption, the time-$0$ local model is calibrated to the binary option prices (which precisely means that $\esp{\indi{i^0_\ell=-1}}=P_0(\ell)$ holds for all $0 \le \ell \le T$).
{\rm\hfill\break \textbf{(iii)}} Let $\ellmax < \ell \le T$. We show that $\{i^0_\ell=1\}\cap\{\ell \le \theta^0\} = \emptyset$, which implies by \eqref{e:ab} that $b_0(\ell)=0$. If $i^0_\ell=1$, then $i^0_0=\cdots=i^0_\ell=1$ (as $-1$ is an absorbing state in the local models). In particular, since $\ellmax<\ell$, $q^0(\ellmax,i^0_{\ellmax}) = q^0(\ellmax,1)=0$ by definition of $\ellmax$, meaning that $\theta^0\le\ellmax<\ell$. This proves, as required, that $\{i^0_\ell=1\}\cap\{\ell \le \theta^0\} = \emptyset$.
{\rm\hfill\break \textbf{(iv)}} We last show, for $\ellmax<\ell\le T$, that $\{i^0_\ell=-1\}\cap\{\theta^0\ge\ell\}=\{i^0_{\ellmax}=-1\}$, which implies by \eqref{e:ab} that $a_0(\ell)=\frac{\bQ(i^0_{\ellmax}=-1)}{P_0(\ell)}$, and the proof is concluded by invoking that $\bQ(i^0_{\ellmax}=-1)=P_0(\ellmax)$ as the time-$0$ local model is calibrated to the binary options prices.\\
First, if $i^0_{\ellmax}=-1$, then, for all $\ellmax \le k \le T$, $i^0_k = -1$ and $q^0_k = q^0(k,i^0_k)=q^0(k,-1)=T-k$ as $-1$ is an absorbing state in the local model. In particular, we have $i^0_\ell=-1$. In addition, we proved $\theta^0 \ge \ellmax$, which implies $q^0_k \neq 0$ for all $k < \ellmax$. Besides, $q^0_k=T-k>0$ for all $\ellmax \le k < T$. In conclusion, $q^0_k\neq 0 $ for all $0 \le k < T$, hence $\theta^0=T \ge \ell$. This proves $\{i^0_{\ellmax}=-1\} \subset \{i^0_\ell=-1\}\cap\{\theta^0\ge\ell\}$. \\ Conversely, if $i^0_\ell=-1$ and $\theta^0 \ge \ell$, since by assumption
$\ellmax<\ell$ also holds throughout this part (iv) of the proof, therefore $\ellmax<\ell\le\theta^0$, hence $q^0_{\ellmax}>0$. Since $0<q^0_{\ellmax} \in \{q^0(\ellmax,1),q^0(\ellmax,-1)\}$ and $q^0(\ellmax,1)=0$ by definition, one necessarily has $q^0_{\ellmax}=q^0(\ellmax,-1)$ and hence $i^0_{\ellmax}=-1$.~\finproof 
 
\brem To obtain such simple formulas for $a_0(\ell)$ and $b_0(\ell)$, $0 \le \ell \le T$, we heavily make use of the fact that, in the time-$0$ calibrated local model, the state $-1$ is absorbing, which implies that $q^0(\ell,-1)=T-\ell$. This is not valid in the fair valuation model, preventing us from providing simple formulas for the hedging ratios $A_0(\ell)$ and $B_0(\ell)$. However, these can still be computed  exactly, the way explained in Section  \ref{ssse:no events} below (see in particular Lemma \ref{lem:exact no}).
\erem
 
Hereafter, whenever a random variable $\xi$ is constant on an event $A$, with a slight abuse of notation, we denote its value on $A$ by $\xi(A)$. 

\subsection{Bad Trader's XVAs} \label{ssse:wo events}

    In this section, we study how to compute the various stochastic processes introduced in Section \ref{s:call} regarding a bad trader of Definition \ref{ex: bad trader}, buying the range accrual studied in Section \ref{s:price hedge} and statically hedging it as postulated in Assumption \ref{e:hypcont}.

    Since the bad trader calls back the asset no later than the model switch time $\ts$ (see \eqref{eq:optimal stopping}), by \eqref{e:thetausdis}, the only relevant events in his case are the following partition of $\Omega$:
    \beql{e:events wo}
    & \Awo_{\Td +1} %
    =\left\{ I^{}_0=1,\dots,I^{}_\Td=1 \right\} \mbox{ and, for } 1 \le l \le\Td , \\
    &\Awo_l   
    =
    \{I^{}_0=1,\dots,I^{}_{l-1}=1,I^{}_l=-1\},
    \eeql
where $\Awo_{\Td +1}$ corresponds to the extreme event never occurring on $0\, ..\,\Tb$, while, for $k < l \le\Td $, $\Awo_l$ corresponds to the extreme event first occurring at time $l$ (assuming $I_0=1$). Note that $\Awo_l$ is $\F_l$ measurable, for each $1 \le l \le\Td ,$ while $\Awo_{\Td +1}$ is $\F_\Td$ measurable. 
For  $l \le\Td +1 $ and $k\le l\wedge T$, $I_k(\Awo_l)$ is obviously given by
    $I_k(\Awo_l) = \mathbf{1}_{k<l}-\mathbf{1}_{k=l}.$ 
The stopping times $\tau_s$ and $\texbad$ are also constant on each $\Omega_l$, $1 \le l \le T+1$. Namely, \eqref{e:thetausdis} and \eqref{eq:optimal stopping} imply, for all $1 \le l \le T+1$:
     \beql{e:thetaudiscrs}
     & \tau_s(\Awo_l)=\inf \{k; I_k=-1\}\wedge\Td = l\wedge T,\\
     &  \tau^{bad}_e (\Awo_l)=  \inf \left\{ k <  l\wedge T  ;\, q^{k}(k,1)=0 \right\}\wedge( l\wedge T)   ,
     \eeql
     which can be determined from the $q^k(k,1)$, $1 \le k \le T$, computed via \eqref{e:qd}.
Moreover, with the notations \eqref{e:uv} at hand,  
as proved in Section \ref{ss:B1}:
    \bl \label{l:Rwo} For every $ k \le\Td $ and $1\le  l \le\Td +1$, the $\F_k$ conditional probabilities of the partitioning events $\Awo_\lambda,$ $1\le \lambda\le\Td +1$, are constant on each $\Awo_l$, where they are worth
    \beql{r:rkll} 
      & \probc{k}{\Awo_\lambda}(\Awo_l)= \un_{  k \ge \lambda}\un_{l=\lambda}+  \un_{k < \lambda} \un_{l>k}\left(\prod_{m=k+1}^{\lambda-1} u_{m}  \right)v_\lambda   \sp 1\le  \lambda \le\Td , \mbox{  and }\\
    & \probc{k}{\Awo_{\Td +1}}(\Awo_l)  = 
    \un_{l>k}\prod_{m=k+1}^{\Td} u_m .
    \eeql 
    \el
       
Since the market is represented by the process $I$ and the processes relative to the bad trader are all stopped at $\tau^{bad}_e$, the corresponding study boils down to understanding computations relative to $\F_{\tau^{bad}_e} \cap \sigma(I_k, k\le T)$ measurable random variables. Now, for such random variable, the following properties are proved in Section  \ref{ss:B1}:
    
    \bl \label{lem:exact wo} {\rm \textbf{(i)}} Let %
    $\xi%
    $ be
    an $\F_{\tau^{bad}_e}\cap\sigma(I_k,k\le T)$ measurable random variable.
     Then, $\xi $ is constant on each $\Awo_l, 1 \le l \le\Td  +1 $. 
    {\rm\hfill\break \textbf{(ii)}} Let $\xi$ be a random variable constant on each $\Awo_l$, $1 \le l \le T+1$. Then, for each $0 \le k \le\Td $ and $1 \le l \le\Td  + 1$, $\espc{k}{\xi}$, $\VaR_k(\xi)$ and $\ES_k(\xi)$ are constant on $\Awo_l$; in particular,
    \beql{e:tauemes2}
    &\espc{k}{\xi}(\Awo_l)
    =  \sum_{\lambda=1}^{\Td+1}
    \xi(\Awo_\lambda) \probc{k}{\Awo_\lambda}(\Awo_l).
    \eeql
If in addition $l \le k$, then \beql{e:ekxo}&\espc{k}{\xi}(\Awo_l)=\VaR_k(\xi)(\Awo_l)=\ES_k(\xi)(\Awo_l)=\xi(\Awo_l). 
\eeql
    \el

    We now apply the above lemmas to the random variables associated with the bad trader.
    
    \bp\label{p:bad}
    Let Assumption \ref{e:hypcont} be in force. Let $k\le\Td $.
    {\rm\hfill\break \textbf{(i)}} We have 
    \beql{e:badter} 
    &\pal^{bad}_{\tk} = %
    \cQ_{\tk\wedge\texbad}+\theq_{\tk\wedge\texbad}\indi{\tk\wedge\texbad<\ts}+\theQ_{\tk\wedge\texbad}\indi{\tk\wedge\texbad=\ts} - (\cP^{loc}_{\tk\wedge\texbad}+ \theP^{loc}_{\tk\wedge\texbad})%
    \\ &\quad\quad\quad-%
    \indi{k\ge\texbad}\indi{\texbad=\ts}\theQ_{\texbad}%
    ,\\
    & \HVA^{bad}_{\tk} = \underbrace{ \left( \theq _{\tk\wedge \texbad}-\theQ _{\tk\wedge \texbad} \right)\indi{\tk\wedge \texbad<\ts}}_{=:{U^{bad}_k}} +\underbrace{ \espc{k}{   \theQ_{\texbad}\indi{\texbad< \ts }}}_{=:{V^{bad}_k}%
    }\\&\quad\quad\quad+  \indi{\tk<\texbad} \underbrace{\espc{k}{\indi{\texbad=\ts}\theQ_{\texbad} } }_{ =:{W^{bad}_k}%
    }+\underbrace{ \cQ_{\tk\wedge \texbad}+\theQ_{\tk\wedge \texbad} -\espc{k}{  \cQ_{\texbad} +\theQ_{\texbad} }}_{va(K^{\texbad})_k%
    }. 
    \eeql 
      {\rm\hfill\break \textbf{(ii)}} The random variables 
    $ \pal^{bad}_{\tk}$ and $\HVA^{bad}_{\tk}$ are constant on each of the $\Awo_l$, where their values can be computed using Propositions \ref{prop: pricing hedging fair model}-\ref{prop: pricing hedging local model} and Lemmas \ref{l:Rwo}-\ref{lem:exact wo}.
    {\rm\hfill\break \textbf{(iii)}} 
    $\EC^{bad}_k$, as defined in \eqref{e:eckvacont} specified to the bad trader dealing the range accrual, is constant on each of the $\Awo_l$, 
    with $\EC^{bad}_k (\Awo_l)=0$ for $l \le k$ and a constant independent of $l$, denoted by $\EC^{bad}(k)$ and also computable by Propositions \ref{prop: pricing hedging fair model}-\ref{prop: pricing hedging local model} and Lemmas \ref{l:Rwo}-\ref{lem:exact wo}, for $l > k$.
    {\rm\hfill\break \textbf{(iv)}} $\KVA^{bad}_k$, as defined in \eqref{e:eckvacont} specified to the bad trader dealing the range accrual, is constant on each $\Awo_l$, $1 \le l \le T+1$. In particular, we have $\KVA^{bad}_k(\Omega_l)=0, 1 \le l \le k \le T$, and
    \beql{e:badkva} 
    \KVA^{bad}_0 
    &=\hurdle\sum_{k=0}^{\Td-1}e^{-\hurdle k}%
    \sum_{\lambda=k+1}^{\Td+1} \max(\EC^{bad}(k), \KVA^{bad}_k(\Awo_\lambda)) \R[\Awo_\lambda]
     .
    \eeql
    \ep
    
     \proof We fix $0 \le k \le T$.\\
     {\rm \textbf{(i)}} The equations for $\pal^{bad}_k$ and $\HVA^{bad}_k$ follow from \eqref{e:bad pnl} and \eqref{e:HVA gen}, recalling that $P^{loc}_0=q_0$ (see after Assumption \ref{e:hypcont}) and $\indi{\texbad<\ts} q_{\texbad}=0$ (by \eqref{eq:optimal stopping} and \eqref{eq:theta star}).
     {\rm\hfill\break \textbf{(ii)}} For each $0 \le k \le T$, the random variables
     \beql{variables} &\pnl^{bad}_k, U^{bad}_k, \indi{\texbad<\ts}Q_{\texbad}, \\ &\indi{k<\texbad}\indi{\texbad\ge\ts}Q_{\texbad}, \text{ and } \cQ_{k\wedge\texbad}+Q_{k\wedge\texbad}-(\cQ_{\texbad}+Q_{\texbad})\eeql are obviously $\F_{\texbad}$ measurable. From \eqref{e:thetaudiscrs}, $\ts= \inf\{k;\, I_k=-1\}$ is $\sigma(I_k,k\le T)$ measurable, and so is $\texbad = \inf\{k;\, q^k(k,1)=0\}\wedge\ts$, as $ \inf\{k;\, q^k(k,1)=0\}$ is deterministic. By definition \eqref{e:rangeaccr disc} and \eqref{l:epT discbisprel}, the processes $\cQ$ and $\cP^{bad}$ are $\sigma(I_k,k\le T)$ adapted. By Propositions \ref{prop: pricing hedging fair model} and \ref{prop: pricing hedging local model}, the processes $q = (q^k(k,I_k))_{0\le k \le T}$ and $Q = (Q^k(k,I_k))_{0 \le k \le T}$ are also $\sigma(I_k,k\le T)$ adapted. So is also $P^{bad}$, by  \eqref{e:Plocex} and \eqref{l:epT disc}. Hence all the random variables in \eqref{variables} are $\F_{\texbad}\cap\sigma(I_k,k\le T)$ measurable. By Lemma \ref{l:Rwo}(i), they are therefore constant on each of the $\Omega_l$, $1 \le l \le T+1$. By Lemma \ref{l:Rwo}(ii), this then implies that $V^{bad}_k = \espc{k}{\indi{\texbad<\ts}Q_{\texbad}}$, $\indi{k<\texbad}W_k = \espc{k}{\indi{k<\texbad=\ts}Q_{\texbad}}$ and $va(K^{\texbad})_k = \espc{k}{\cQ_{k\wedge\texbad}+Q_{k\wedge\texbad}-(\cQ_{\texbad}+Q_{\texbad})}$ are constant on each $\Omega_l$, $1 \le l \le T+1$, and so is in turn $\HVA^{bad}_k$. 
     {\rm\hfill\break \textbf{(iii)}} $\EC^{bad}_k$ is, by \eqref{e:eckvacont}, the $\F_k$ conditional expected shortfall of a random variable which is, by (ii), constant on each $\Awo_l$, $1 \le l \le T+1$. By Lemma \ref{l:Rwo}(ii), $\EC^{bad}_k$ is also constant on each $\Awo_l$. 
  Moreover, if $l\le k$, \eqref{e:ekxo} shows that $\EC^{bad}_k(\Awo_l) = -(pnl^{bad}_{(k+1)\wedge T}(\Awo_l)-pnl^{bad}_k(\Awo_l))+\HVA^{bad}_{(k+1)\wedge T}(\Awo_l)-\HVA^{bad}_k(\Awo_l)$. But this is equal to $0$ as the processes $pnl^{bad}$ and $\HVA^{bad}$ are stopped at $\texbad$ (see \eqref{e:badter}), which is $\le l$  on $\Awo_l$ (see \eqref{e:thetaudiscrs}).
    Moreover, the first line of
    \eqref{r:rkll} shows that $\R_k[\Awo_\lambda](\Awo_l)$ is equal to 0 for $l\le k$ and does not depend on $l$ for $l>k$, 
    which implies
    the last statement regarding EC.
    {\rm\hfill\break \textbf{(iv)}} By backward induction on $k$, $\KVA^{bad}_k$ is  constant on each $\Awo_l$, $1 \le l \le T+1$. In fact, $\KVA^{bad}_T= 0$, while assuming the induction hypothesis at rank $k+1$ yields by \eqref{e:eckvacont} that $\KVA^{bad}_k$ is  the $\F_k$ conditional expectation of a random variable which is, by (iii), constant on each $\Awo_l$, $1 \le l \le T+1$. Hence, by Lemma \ref{lem:exact wo}(ii), $\KVA^{bad}_k$ is also constant on each $\Awo_l$, $1 \le l \le T+1$. In addition, we have, by \eqref{e:eckvacont} (in discrete time), for $1 \le l \le k \le T$,
    \bel
    &\KVA^{bad}_k(\Awo_l)=\hurdle \espc{k}{ \sum_{t =k+1}^{\Td} e^{- \hurdle (t-k) } \max(\KVA^{bad}_t,\EC^{bad}_t) }(\Awo_l) \\
    &=\hurdle \sum_{\lambda=1}^{\Td+1} \left(\sum_{t =k+1}^{\Td} e^{-\hurdle (t-k)}\max(\KVA^{bad}_t(\Awo_\lambda),\EC^{bad}_t(\Awo_\lambda))\right)\R_k[\Awo_\lambda](\Awo_l),
    \eel
    by \eqref{e:tauemes2}.
    By \eqref{r:rkll}, since $l \le k$, $\R_k[\Awo_\lambda](\Awo_l) = \mathbf{1}_{\lambda=l}$. Hence
    \bel
    &\KVA^{bad}_k(\Awo_l) = \hurdle \sum_{t =k+1}^{\Td} e^{-\hurdle (t-k)}\max(\KVA^{bad}_t(\Awo_l),\EC^{bad}_t(\Awo_l)).
    \eel
    By (iii), we have ${\rm EC}^{bad}_t(\Omega_l) = 0$ as $l \le k < t$, hence
    \bel
    &\KVA^{bad}_k(\Awo_l) = \hurdle \sum_{t =k+1}^{\Td} e^{-\hurdle (t-k)}\big(\KVA^{bad}_t(\Awo_l)\big)^+,
    \eel
    and a straightforward backward induction in $k$, starting from $\KVA^{bad}_T(\Awo_l) = 0$, shows that $\KVA^{bad}_k(\Awo_l) = 0$ for $k \ge l$. 
    
    Besides, \eqref{e:HVA gen} yields
    \bel
    &\KVA^{bad}_0=\hurdle  \esp{ \sum_{l =1}^{\Td} e^{- \hurdle l } \max(\KVA^{bad}_l,\EC^{bad}_l)}\\& =\hurdle   \sum_{l =1}^{\Td}   e^{-\hurdle l}  \sum_{\lambda=1}^{\Td+1}
    \max(\KVA^{bad}_l(\Awo_\lambda),\EC^{bad}_l(\Awo_\lambda))\R [\Awo_\lambda],
    \eel
    and since $\KVA^{bad}_l(\Awo_\lambda) = \EC^{bad}_l(\Awo_\lambda) = 0$ for $\lambda \le l$, we obtain \eqref{e:badkva}.~\finproof

    \subsection{Not-So-Bad Trader's XVAs} \label{ssse:no events}
    
We now perform the computations regarding the not-so-bad trader of Definition \ref{ex: nsb trader}, buying the range accrual and statically hedging it along Assumption \ref{e:hypcont}.

    To ease the study, we make the following assumption (which will be satisfied in our numerics).
    \bhyp \label{ass: Q 1 zero}
    For all $0 \le k \le\Td $, we have $Q^{}(k,1)=0$.
    \ehyp
\noindent Then, starting from $I_0=1$, in the global model, it would be optimal for the bank to call the asset immediately, see \eqref{e:tau t} with $t=0$. But the use of the local model may lead the trader to overvalue the claim and to a delayed exercise decision. 
    \brem
    Playing with different numerical parametrizations of the model often leads to $Q(\cdot,1)\equiv0$.
    In particular, for any positive parameter
    $\gamma_{\Td-1} $, 
    forcing $Q (\cdot, 1)=0$ and the continuation value $ - \h e^{-2\h\gamma^{}_k}  + u_{k+1}  Q^{}(k+1,1) +v_{k+1}  Q^{}(k+1,-1)$  to be 0 in the equation for $ Q(k,1)$ in \eqref{e:Qd} yields $Q  (\Td,\cdot)=0$ and, for decreasing $k\le\Td -1$,
    \bel 
    & Q(k,-1)
    =  \h e^{-2 \h\gamma_k}   + \frac12\left(1 + e^{-2\h\gamma_k}\right)Q(k+1,-1) , \\
    & 1%
    = \frac12 \left(  e^{ 2\h\gamma_{k-1}}-1\right) Q(k ,-1)\mbox{ i.e.~} \gamma_{k-1}=\frac 1{2%
    } \ln\big(1+\frac {2%
    } { Q(k ,-1)}\big),
    \eel 
    which iteratively determine
    $Q(k,-1)> 0$ and $\gamma_k>0$.  
    This provides a whole family of models for which $Q(\cdot,1)\equiv 0$ (i.e.\ Assumption \ref{ass: Q 1 zero} holds), parameterized by 
    $\gamma_{\Td-1} >0$.
    \erem
    Under Assumption \ref{ass: Q 1 zero}, the only events that are relevant to the not-so-bad trader are the following partition of $\Omega:$ 
    \beql{e:events no}
    &\Ano_{\Td +1,\Td +1} = \{I^{}_0=1,\dots,I^{}_\Td=1\},\\
    &\Ano_{l,\Td +1} = \{I^{}_0=1,\dots,I^{}_{l-1}=1,I^{}_l=-1,\dots,I^{}_{\Td}=-1\}, \quad 1 \le l \le\Td , \\ 
    &\Ano_{l,m}  = \{I^{}_0=1,\dots,I^{}_{l-1}=1,I^{}_l=-1,\dots,I^{}_{m-1}=-1,I^{}_m=1\}, \quad 1 \le l < m \le\Td , 
    \eeql
    where $\Ano_{\Td +1,\Td +1}$ corresponds to the extreme event never occurring on $0\, ..\, \Tb$; for $1\le l\le T,$ $\Ano_{l,\Td +1}$ corresponds to the extreme event first happening at time $l$ and never ceasing on $l\, ..\, \Tb$; for $1 \le l < m \le\Td $, $\Ano_{l,m}$ corresponds to the extreme event first occurring at time $l$ and then first ceasing at time $m$. Note that $\Ano_{l,m}$ is $\F_m$ measurable, for each $1 \le l<m \le\Td ,$ and $\Ano_{l,\Td +1}$ is $\F_\Td$ measurable, for $1 \le l \le\Td +1$.
    
    The index-set of these market events is
    \bel\cI := \left\{( l , m );\, 1\le l < m \le\Td \right\}\cup \left\{( l ,\Td +1 );\,1\le l   \le\Td  +1\right\}.
    \eel
For $(l,m) \in \cI$, we have $\Ano_{l,m} \subset \Awo_l$ (compare \eqref{e:events no} and \eqref{e:events wo}), $I_k(\Ano_{l,m}) =  \mathbf{1}_{k<l}-\mathbf{1}_{l \le k < m}+\mathbf{1}_{k=m}$ for $k \le m \wedge T$, $\ts(\Ano_{l,m})=l \wedge T$, and
    \bel
    \texnsb(\Ano_{l,m}) = \texbad(\Ano_{l,m}) \indi{\texbad(\Ano_{l,m}) < l \wedge T} + \inf \left\{ k \ge \ts(\Ano_{l,m});\, Q_k = 0\right\} \indi{\texbad(\Ano_{l,m}) = l \wedge T}.
    \eel
Also note that $\texnsb(\Ano_{l,m}) \le m$ if $m \le T$ as, on $\{\texbad(\Ano_{l,m}) = l\}$, $Q_m = Q(m,I_m(\Ano_{l,m})) = Q(m,1) = 0$ by Assumption \ref{ass: Q 1 zero}. 
 Moreover, with the notations \eqref{e:uv} at hand, 
    as proved in Section \ref{ss:B2}:
    \bl \label{l:rkllno} For every $0\le k\le\Td $,
    the $\F_k$ conditional probabilities of the partitioning events $\Ano_{\lambda, \mu}$, $(\lambda,\mu)\in \cI,$ are
constant on each $ \Ano_{l, m}$, $(l,m)\in \cI $, where they are worth
    \beql{r:rkllno} 
     & \probc{k}{\Ano_{\lambda,\mu }}(\Ano_{l,m})
    =  \big(\un_{k<l\wedge \lambda }+\un_{k\ge  l\wedge \lambda }\un_{l= \lambda }(\un_{ k<m\wedge \mu }+\un_{ k\ge m\wedge \mu }\un_{m=\mu })\big)\times\\&\qqq
    \Big(\un_{ k\ge\mu }+
    \un_{\lambda\le  k<\mu } 
    \big( \prod_{r=k+1 }^{\mu-1 } \even_r\big)
     \odd_{\mu}  +\\&\qqq\qqq \un_{ k<\lambda } 
    \big( \prod_{r=k+1 }^{\lambda-1 } \even_{r}\big)
     \odd_{\lambda}
     \big(\prod_{r=\lambda+1 }^{\mu-1} \even_{r}\big) \odd_{\mu }\Big)  
     \sp 1\le \lambda<\mu \le\Td ,  \\
    &\probc{k}{\Ano_{\lambda,\Td +1}}(\Ano_{l,m})  =
    \big(\un_{k<l\wedge \lambda }+\un_{k\ge  l\wedge \lambda }\un_{l= \lambda }\un_{ k<m }\big)\times\\&\qqq
    \Big(\un_{ k\ge\lambda } 
     \prod_{r=k+1 }^{\Td } \even_{r}+\un_{ k<\lambda } 
    \big( \prod_{r=k+1 }^{\lambda-1 } \even_{r}\big)
     \odd_{\lambda}
     \big(\prod_{r=\lambda+1 }^{\Td } \even_{r}\big)\Big) ,
     1\le \lambda \le\Td  ,
     \mbox{ and }
     \\
     & \probc{k}{\Ano_{\Td +1,\Td +1}}(\Ano_{l,m}) =
     \un_{k<l}\prod_{r=k+1 }^{\Td } \even_{r} .
    \eeql 
    \el

   Since the market is represented by the process $I$ and the processes related to the not-so-bad trader are all stopped at $\texnsb$, the study regarding the latter boils down to understanding computations relative to $\F_{\texnsb} \cap \sigma(I_k, k\le T)$ measurable random variables. Now, as proved in Section \ref{ss:B2}:
    \bl \label{lem:exact no}   {\rm \textbf{(i)}} Let $\xi$ be an $\F_{\texnsb} \cap \sigma(I_k, K \le T)$ measurable random variable. Then $\xi$ is constant on each $\Ano_{l,m}$, $(l,m)\in\cI$.
    {\rm\hfill\break \textbf{(ii)}}   Let $\xi$ be a random variable constant on each $\Ano_{l,m}$, $(l,m)\in\cI$. Then, for each $0 \le k \le\Td $ and $(l,m)\in\cI$, $\espc{k}{\xi}$, $\VaR_k(\xi)$ and $\ES_k(\xi)$ are  constant on each $\Ano_{l,m}$; in particular,
    \beql{e:tauemes2no}
    & 
    \espc{k}{\xi}(\Ano_{l,m}) =%
    \sum_{(\lambda,\mu)\in\cI}  
    \xi(\Ano_{\lambda,\mu}) \probc{k}{\Ano_{\lambda,\mu}} (\Ano_{l,m}). 
    \eeql 
    \el
    
    We now apply these abstract results to the random variables associated with the not-so-bad trader. The proof is similar to the proof of Proposition \ref{p:bad} and thus omitted.

    \bp\label{p:notsobad}
    Let $k\le\Td $.
    {\rm\hfill\break \textbf{(i)}} We have 
    \beql{e:noter} 
    &  \pal^{nsb}_{\tk} = %
    \cQ_{\tk \wedge \texnsb} + \theq_{\tk \wedge \texnsb}\indi{k\wedge\texnsb<\ts}+\theQ_{\tk \wedge \texnsb}\indi{\tk\wedge\texnsb\ge\ts} \\ &\quad\quad\quad\quad - \big(\cP^{loc}_{\tk \wedge \texnsb \wedge \ts}+  P^{nsb}_{\tk \wedge \texnsb \wedge \ts}\big) - \left(\cP^{fair}_{\tk \wedge \texnsb}-\cP^{fair}_{\ts}+P^{fair}_{\tk \wedge \texnsb}-P^{fair}_{\ts}\right)\indi{\tk \wedge \texnsb \ge \ts}
    ,\\
    & \HVA^{nsb}_{\tk} = \underbrace{\indi{\tk \wedge \texnsb < \ts}( \theq_{\tk \wedge \texnsb}-\theQ_{\tk \wedge \texnsb} )}_{=: {U^{nsb}_k }} 
    + \underbrace{ \cQ_{k \wedge \texnsb}+\theQ_{k \wedge \texnsb} -\espc{k}{\cQ_{\texnsb}+\theQ_{\texnsb}}}_{=va(K^{\texnsb})_k%
    }  
    .\eeql
    {\rm\hfill\break \textbf{(ii)}}
    The random variables $\pal^{nsb}_{\tk}$ and $\HVA^{nsb}_{\tk} $ are constant on each of the $\Ano_{l,m}$, $(l,m)\in\cI,$ where their values can be computed by application of Propositions \ref{prop: pricing hedging fair model}-\ref{prop: pricing hedging local model} and Lemmas
    \ref{l:rkllno}-\ref{lem:exact no}.  
    {\rm\hfill\break \textbf{(iii)}}
    $\EC^{nsb}_k$, as defined in \eqref{e:eckvacont} specified to the not-so-bad trader dealing the range accrual, %
    is constant on each of the $\Ano_{l,m}$, $(l,m)\in\cI$.
    {\rm\hfill\break \textbf{(iv)}} $\KVA^{nsb}_k$, as defined in \eqref{e:eckvacont} specified to the not-so-bad trader dealing the range accrual, is constant on each of the $\Ano_{l,m}$, $(l,m)\in\cI$. In addition, 
 \bel
    &\KVA^{nsb}_0 
    =\hurdle   \sum_{l=0}^{\Td-1}  e^{-\hurdle l} \sum_{(\lambda,\mu)\in\cI}
    \max(\KVA^{nsb}_l(\Og_{\lambda,\mu}),\EC^{nsb}_l(\Og_{\lambda,\mu})) \R [\Awo_{\lambda,\mu}]  .
    \eel  
    \ep

    \brem\label{rem: no W} In the $\HVA^{nsb}$ equation in \eqref{e:noter}, we see no 
$V^{nsb}_k :=\espc{k}{  \theQ_{\texnsb}\indi{\texnsb< \ts }}$ analog of the $V^{bad}_k$ term in the $\HVA^{bad}$ equation \eqref{e:badter}. This is because,
on $\{\texnsb< \ts \}, $ $\theQ_{\texnsb}$ vanishes by Assumption \ref{ass: Q 1 zero}, hence $V^{nsb}_k=0$.
    We see no
    $W^{nsb}_k := \espc{k}{\indi{\texnsb\ge\ts}Q_{\texnsb}}$ analog of the $W^{bad}_k$ term either because, on $\{\texnsb \ge \ts\}$, $Q_{\texnsb}=0$ holds by Definition \ref{ex: nsb trader}, hence $W^{nsb}_k=0$.
    \erem

           Corollary \ref{cor:calib} allows one to recalibrate the local model analytically conditionally on any scenario of the fair valuation model.
            Proposition \ref{prop: pricing hedging local model} allows one to price analytically  in the local model shifting along the fair valuation one.  
            Propositions  \ref{p:bad} and \ref{p:notsobad} allow one to compute the HVA and KVA of the bad and of the not-so-bad traders analytically in any scenario of the fair valuation model.
            All in one,  the cost of computing the HVA and the KVA in this setup is reduced to the one of running the exact dynamic programming equations \eqref{e:Qd} for $Q$
           and  \eqref{e:qd} for each $q^k$, $0\le k \le \Td $, along with companion analytical valuations at each nodes of the corresponding computational trees, of sizes O$(T)$ each, hence a total computational cost in 
		O$(T^2)$, and exact computations \b{(in our fully discrete setup we avoid the numerical error inherent to any PDE numerical or Monte Carlo simulation scheme). This simple but representative example illustrates all the ins and outs of recalibration risk and Darwinian model risk, while allowing us to understand how, conversely, such calculations would be unfeasible for a banking portfolio and realistic models: other types of callable assets could in principle be considered following the same logic, with expected similar qualitative insights, but a more complex setup would lead to much more involved computations, with nested numerical optimization for the embedded recalibration task in particular. Not only would this result into an extremely heavy procedure, but it would induce a numerical error obscuring the financial interpretation.}

		\brem \b{The practical relevance of the callable range accrual -- a widely traded asset known to have caused significant losses -- makes it an enlightening example. The probabilistic model that we consider, considering only a finite number of market scenarios, is tailored to the product. It allows one to recalibrate the local model analytically (without any numerical optimization) conditionally on any scenario of the reference model, to compute exactly (i.e.\ without numerical approximations with e.g. PDE methods or Monte-Carlo simulations) all the quantities of interest, and to provide financial interpretations and recommendations.}

\erem
     
    \section{Numerical Results\label{s:num}}
    
    We take $\Tb=10$ years and $\gamma^{}_k \eqdef \int_k^{k+1}  \gamma^{}(s) ds  $
    with \beql{e:gammanum}\gamma^{}(s) := 0.15 - \frac{0.1s}{\Tb}= 0.15 -  0.01 s .\eeql
    Hence $\gamma_k = 0.15 - \frac{0.1}{2T} ((k+1)^2-k^2 )=  0.15 - \frac{0.1}{2T} (2k+1)$, for $0\le k \le \Td -1   $. 
    The jump intensity functions $(\gamma_l)_{l\le \Td-1}$ and %
    $(\nu^0_l)_{l\le \Td-1}$ calibrated to it via \eqref{eq:calib disc} for $k=0$ are represented in Figure \ref{fig:gamma-nu}. 
    
     A nominal (scaling factor) of 100 is applied everywhere to ease the readability of the results.
    Figure \ref{fig:fct-q-Q} displays the pricing functions $Q(t,\mp 1)$ and $q^0(t,\mp 1)
    $ of the callable range accrual 
    in the fair valuation model and in the trader's local model calibrated to it at time 0, computed by the dynamic programming equations of Propositions \ref{prop: pricing hedging fair model}-\ref{prop: pricing hedging local model}.
    The trader's local model overvalues the option, which increases his competitiveness for buying the claim from his client, in line with the first Darwinian principle of Subsection \ref{ss:da}.
    
    \begin{figure}%
    \hspace*{-1cm}
    \begin{subfigure}{.5\linewidth}
      \centering
      \includegraphics[width=\linewidth]{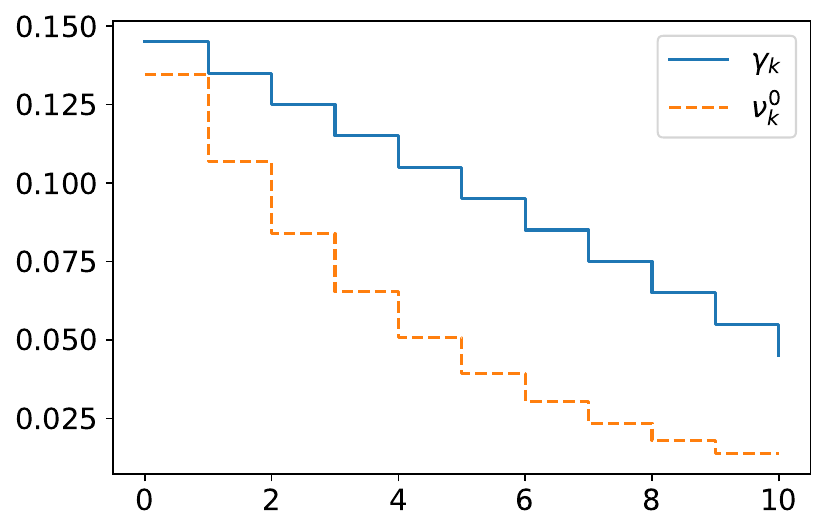}
      \caption{Fair model jump intensity function $(\gamma_k)_{k\le \Td-1}$ and local model jump intensity function $(\nu^0_k)_{k\le \Td-1}$ calibrated to the latter on the binaries at $t=0$.\\
      }
      \label{fig:gamma-nu}
    \end{subfigure}
    \hspace*{0.5cm}
    \begin{subfigure}{.5\linewidth}
      \centering
      \includegraphics[width=\linewidth]{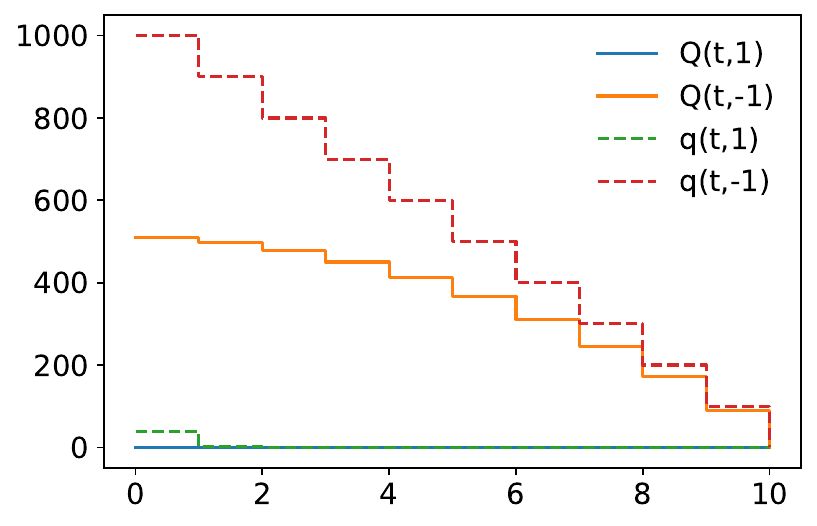}
      \caption{Pricing functions of the range accrual in the global  model,  $Q(\cdot,\cdot)$, and in the local model calibrated to the latter on the binaries at time 0, $q^0(\cdot,\cdot)$.}
      \label{fig:fct-q-Q}
    \end{subfigure}
    
    \hspace*{-1cm}
    \begin{subfigure}{.5\linewidth}
      \centering
      \includegraphics[width=\linewidth]{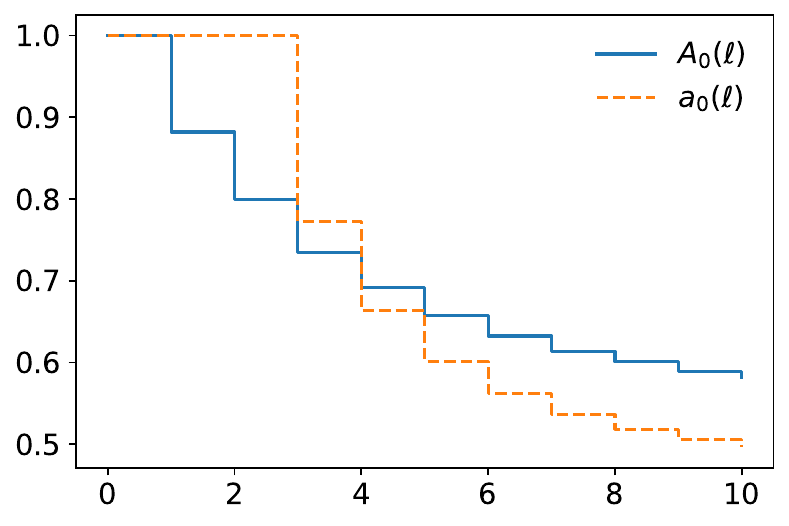}
      \caption{Hedging ratios in the binaries struck along the lower barrier $i=-1$ in the global model, $A_0(\ell)$, and in the local model calibrated to the fair values of all binaries at time 0, $a_0(\ell)$.}
      \label{fig:hedging-A}
    \end{subfigure}
    \hspace*{0.5cm}
    \begin{subfigure}{.5\linewidth}
      \centering
      \includegraphics[width=\linewidth]{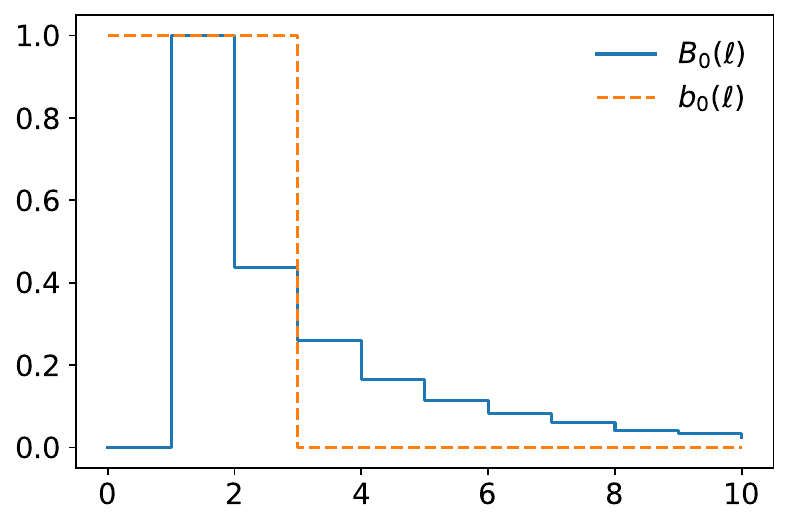}
      \caption{Hedging ratios in the binaries struck along the upper barrier $i=+1$ in the global model, $B_0(\ell)$, and in the local model calibrated to the fair values of all binaries at time 0, $b_0(\ell)$.}
      \label{fig:hedging-B}
    \end{subfigure}
    \caption{Jump intensities, pricing, and greeking functions.}
    \end{figure}

    Note that the pricing function  $Q(\cdot, 1)$ satisfies Assumption \ref{ass: Q 1 zero}. 
    Hence, based on 
    Propositions \ref{p:bad}-\ref{p:notsobad}  and their consequences detailed in Sections \ref{ssse:wo events}-\ref{ssse:no events}, one has numerically access to an exhaustive description of both cases at hand (the bad trader as per Subsection \ref{ssse:wo events}
    and the not-so-bad trader under Assumption \ref{ass: Q 1 zero} as per  Subsection \ref{ssse:no events}), exact within machine precision (only involving discrete dynamic programming equations or exact formulas for path-dependent quantities, without Monte Carlo simulations).

    \subsection{Bad Trader}
    
    For $\gamma(\cdot)$ as per \eqref{e:gammanum}, the dynamic programming equations yield $q^1(1,1)>0$ and $q^2(2,1)=0$. The first equality implies that the trader calls back the option at $t=1$ if and only if the model switch occurs at $t=1$. If $\tau_s>1$, then the trader always calls the asset at $t=2$, whether that $I_2=-1$, i.e.\ $\tau_s=2$, or that $I_2=1$ and, as $q^2(2,1)=0$, it is optimal for the bad trader to exercise.
    Hence the only relevant events are $\Omega_1$, $\Omega_2$ and $\Omega_{T+1=11}$ (on each $\Omega_l$, $l\geq3$, everything happens as on $\Omega_{11}$).
 
    Figure \ref{fig:bad-pnl}, \textit{center panel}, displays $\pnl^{bad}$ on these events.
    We decompose $\pnl^{bad}$ in two terms corresponding to the two lines for $\pal^{bad}_{\tk}$ in \eqref{e:badter}: the cash flows of the first line resulting from holding the option and its hedge plus the corresponding prices (\textit{top panel}) and the ones of the second line accounting for calling the option at zero recovery (\textit{bottom panel}). 
    In the scenarios $\Omega_1$ and $\Omega_2$, where the asset is called due to the model switch, a profit (Figure \ref{fig:bad-pnl}, \textit{top panel}, analyzed in more detail in Section \ref{sse:profit} below), 
    is more than compensated by calling the asset, highly valuable at that moment (Figure \ref{fig:bad-pnl}, \textit{bottom panel}), resulting in an overall loss at the model switch time (Figure \ref{fig:bad-pnl}, \textit{center panel}).

    \begin{figure}
    \hspace{6cm}
    \includegraphics[scale=0.5]{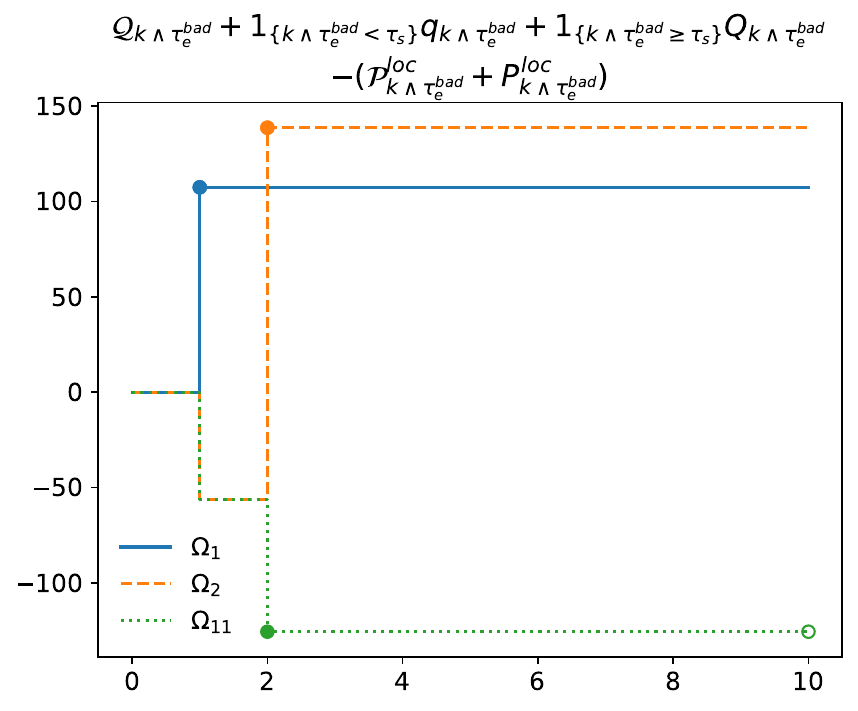}\\
    \includegraphics[scale=0.65]{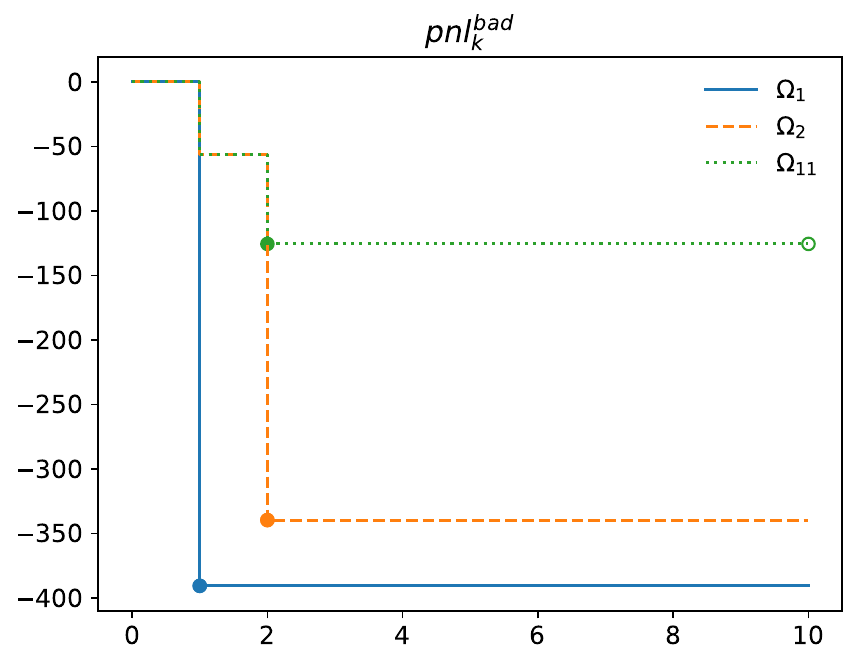}\\
    \hspace*{1cm}
    \hspace{5cm}
    \includegraphics[scale=0.5]{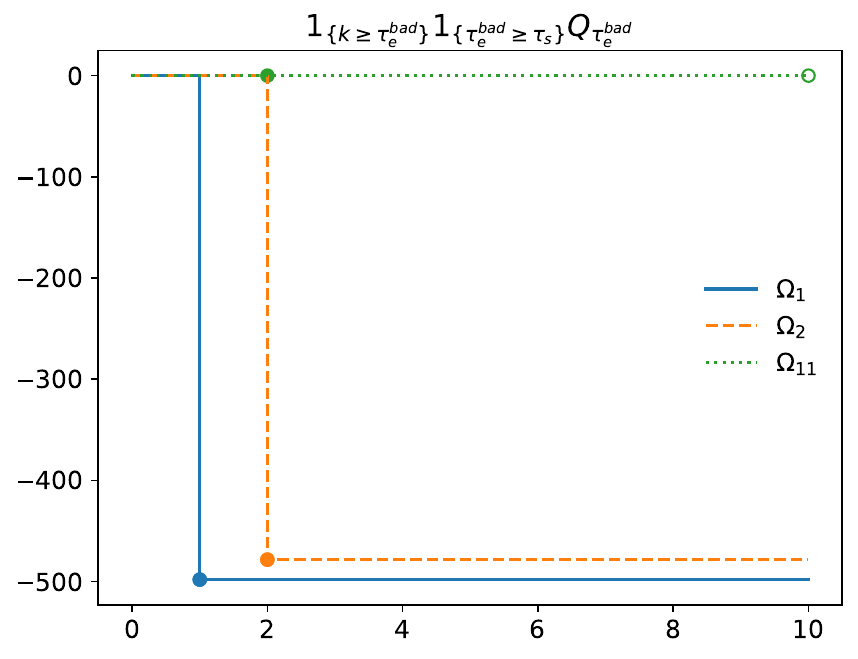}
    \caption{(\textit{center}) bad trader's $\pnl$, (\textit{top}) callable option cash flow and price minus its hedge cash flow and price, (\textit{bottom}) term accounting for calling the product (at zero recovery).}
    \label{fig:bad-pnl}
    \end{figure}

    Figure \ref{fig:bad-hva} displays the process $\HVA^{bad}$ (\textit{top left panel}) and its split into three contributions  (cf.~the  decomposition of $\HVA^{bad}_{\tk}$ in \eqref{e:badter}, see also Remark \ref{rem: interp}): the misvaluation term $U^{bad}_k$ when the trader uses his local model instead of the global one (\textit{top right}), the expected cost of calling the asset at zero recovery %
    $V^{bad}_k + \indi{k<\tau_e^{bad}} W^{bad} _k$ (\textit{bottom left}), and the reserve for suboptimal exercise %
    $va(K^{\texbad})_k$ (\textit{bottom right}). By comparing the top left and right panels, we observe that the HVA on a callable claim can thus be several times greater than the price difference $q-Q$.

    \begin{figure}[ht]
    \hspace*{-1cm}
    \begin{subfigure}{.5\linewidth}
      \centering
      \includegraphics[width=\linewidth]{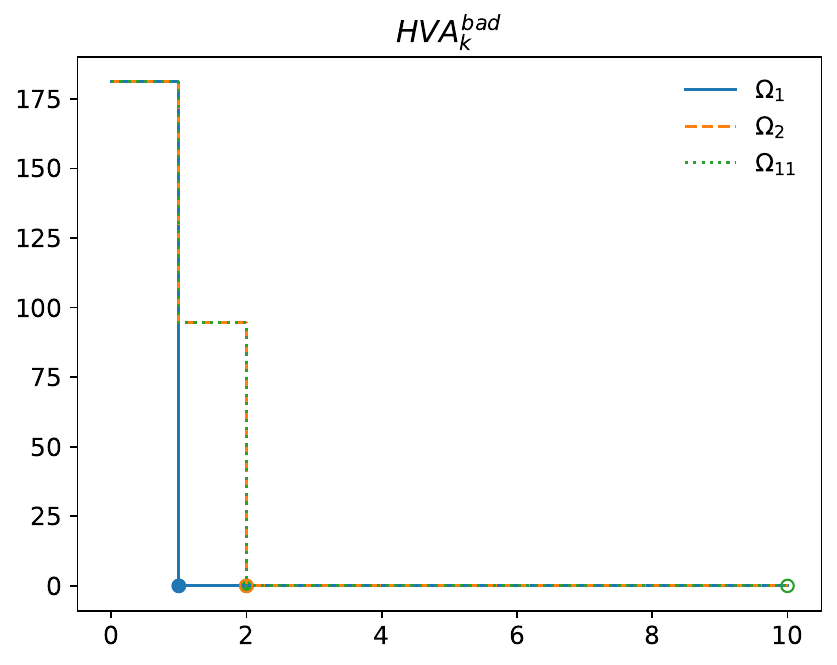}
    \end{subfigure}%
    \hspace*{0.5cm}
    \begin{subfigure}{.5\linewidth}
      \centering
      \includegraphics[width=\linewidth]{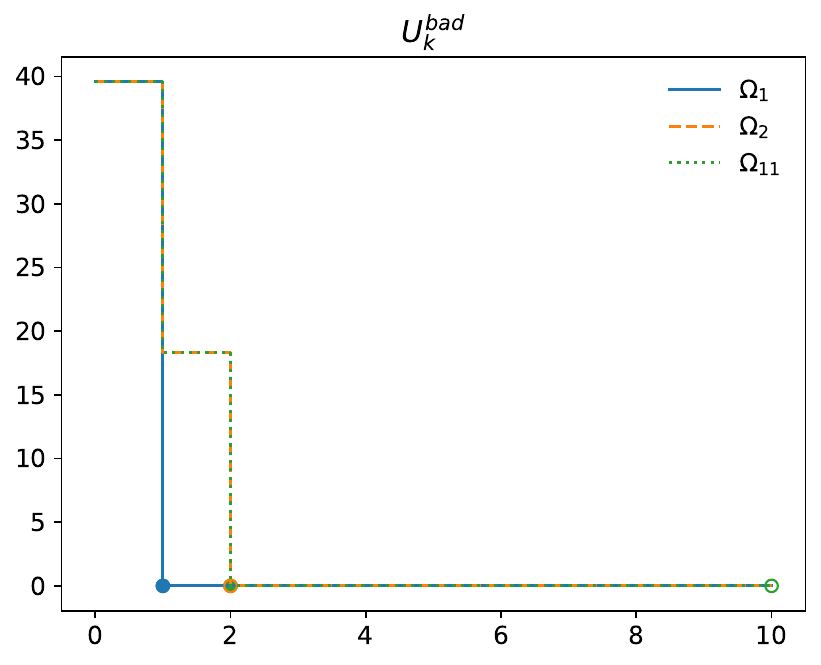}
    \end{subfigure}
    
    \hspace*{-1cm}
    \begin{subfigure}{.5\linewidth}
      \centering
      \includegraphics[width=\linewidth]{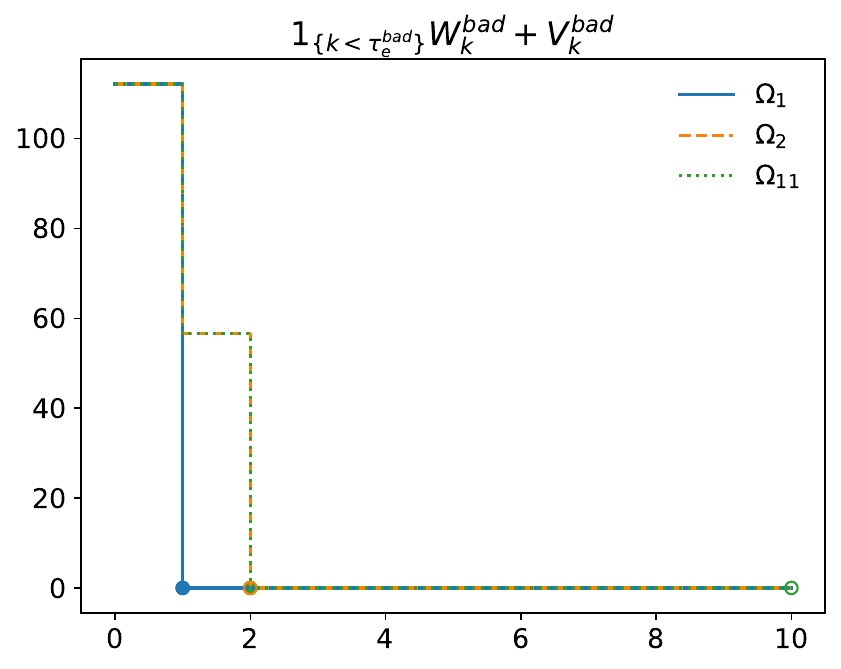}
    \end{subfigure}%
    \hspace*{0.5cm}
    \begin{subfigure}{.5\linewidth}
      \centering
      \includegraphics[width=\linewidth]{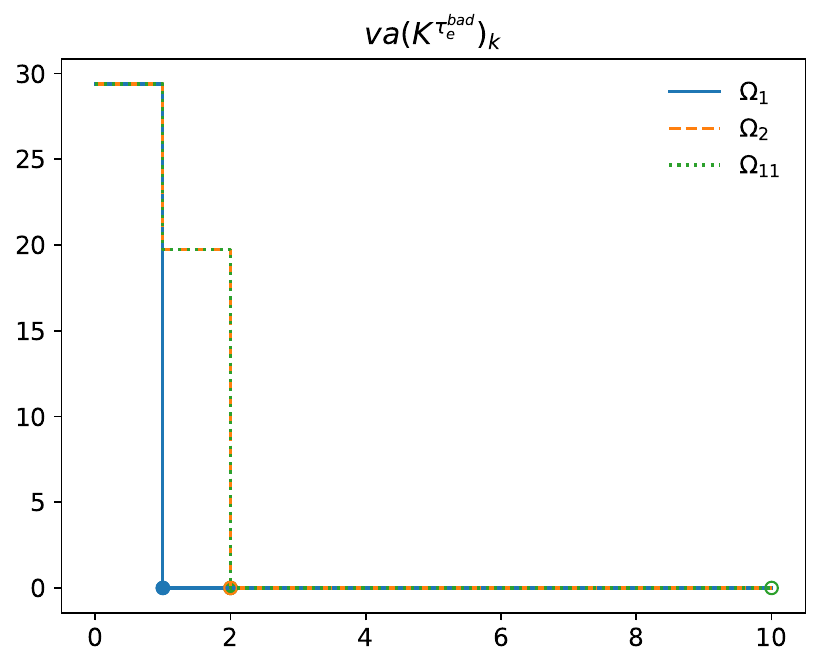}
    \end{subfigure}
     \caption{Bad trader's $\HVA$ and its components. %
    }
    \label{fig:bad-hva}
    \end{figure}
    
    Figure \ref{fig:bad-hva-pnl} displays the HVA compensated pnl process of the \bad trader. We notice that on the event $\Omega_{11}$, where there is no switch and the trader calls back the claim at time 2, the gains resulting from the depreciation of the $\HVA$ cover the $\pal$ losses (the green curve is in the negative), in line with the second Darwinian principle recalled in Subsection \ref{ss:da}.
    But, on $\Omega_1$ and $\Omega_2$, the losses made at $\ts$ supersede the systematic profits made before $\ts$, in line with the third Darwinian principle ofSubsection \ref{ss:da}.
    
    \begin{figure}[H] 
    \centering
    \includegraphics[scale=0.6]{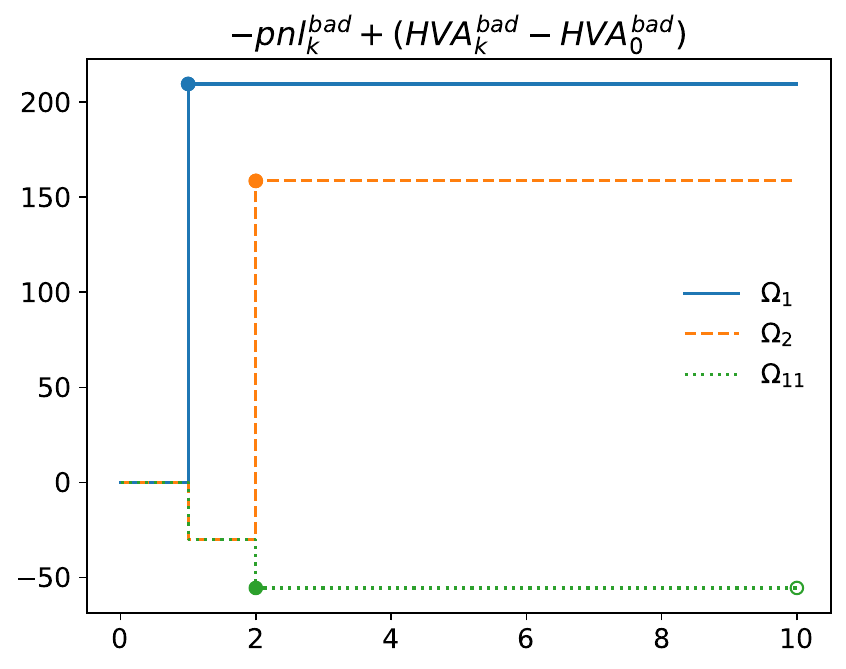} 
    \caption{HVA-compensated loss-and-profits of the bad trader.%
    }
    \label{fig:bad-hva-pnl}
    \end{figure}

    \subsubsection{Detailed understanding of the profit in Figure \ref{fig:bad-pnl}, \textit{top panel} %
    } \label{sse:profit}
   On $\Omega_1 \cup \Omega_2= \{\ts=\texbad\}$, at $\ts$, the bank gets on the asset a cash flow $\cQ_{\ts}-\cQ_{\ts-}=1$, while it pays on the static hedge a cash flow $\cP^{loc}_{\ts}-\cP^{loc}_{\ts-} = a_0(\ts)$, as $I_{\ts}=-1$.
   In addition, in any time-$\ts$ (hence, no longer calibrated) local model and independently of the intensity function $\nu^{\ts}(\cdot)$, as $i^{\ts}_{\ts} = I_{\ts} = -1$ is an absorbing state, at time $\ts$, the asset is worth $q_{\ts}=q^{\ts}_{\ts}=\Tb-\ts$ and the hedge is worth $P^{loc}_{\ts}=\sum_{k=\ts+1}^{\Td} a_0(k)$ (cf.\ \eqref{e:pq} and
 \eqref{e:Plocex}). Hence the profit at the model switch time $\tau_s=1$ or $2$ made before calling the asset, as observed on the top panel of Figure \ref{fig:bad-pnl},
    can be decomposed as follows (see Table \ref{tab:pnl-incr}):
    \beql{e:pnldec} 
    &\left(\cQ_\ts + Q_\ts - (\cP^{loc}_\ts + P^{loc}_\ts)\right) - \left(\cQ_{\ts-1} + q_{\ts-1} - (\cP^{loc}_{\ts-1} + P^{loc}_{\ts-1} \right) =\\
    \nonumber
    & \qq %
    1 + (\Td-\ts)-q^{\ts-1}(\ts-1,1) -  \big( %
    a_0(\ts) + \sum_{k=\ts+1}^{\Td} a_0(k) -P_{\ts-1} \big)  \\
    &\qqq + Q(\ts,-1) - (\Td-\ts) - \big(P^{loc}_{\ts} - \sum_{k=\ts+1}^{\Td} a_0(k)\big),
    \eeql
    where the third line corresponds to the change of valuation model at $\ts$, which is a loss as per Figure \ref{fig:fct-q-Q}. An overall profit (made, at least, \textit{before} calling the asset) means that this loss is more than compensated by a profit coming from the second line, coming from the static hedge not being perfect, especially at $\ts$ (from $\ts$ onward, the perfect hedge would be to short a digital option with payoff $\indi{I_k=-1}$ for each $k> \tau_s$).
    \begin{table}[ht]
    \centering
    \begin{tabular}{|r|r|r|}
        \hline
         &  $\Omega_1$ & $\Omega_2$\\\hline
    $%
    1 + (\Td-\ts)-q^{\ts-1}(\ts-1,1) $ & &  \\
    $- ( %
    a_0(\ts) + \sum_{k=\ts+1}^{\Td} a_0(k) -P^{loc}_{\ts-1})$  & $ 335 $ & $391$ \\\hline
    $ Q(\ts,-1) - (\Td-\ts) - \big(P^{loc}_{\ts} - \sum_{k=\ts+1}^{\Td} a_0(k)\big)$  & $-227$ & $-196$\\\hline
    \end{tabular}
    \caption{The decomposition \eqref{e:pnldec} on the events $\Omega_1$ and $\Omega_2$.}
    \label{tab:pnl-incr}
    \end{table}

    \subsubsection{Numerical stability of the valuation adjustments}\label{ss:numstab}
    
To assess the numerical stability of the proposed metrics, we study the impact on the valuation adjustments $\HVA_0$ and $\KVA_0$ of a parallel shift of the jump intensity $(\gamma_k)_{k \le T-1}$ \eqref{e:gammanum} in the fair valuation model. This also allows us to compute finite-difference approximations of the sensitivities of $\HVA_0$ and $\KVA_0$ with respect to such parallel shifts.

   Namely, given a small (real) shift $s$, we consider the shocked jump intensity  $\gamma^s_k := \gamma_k + s$, $0 \le k \le T-1$ and we compute the corresponding valuation adjustments $\HVA^s_0$ and $\KVA^s_0$ using the numerical procedure described above in the baseline case $s=0$. We then deduce the sensitivities around the baseline case by finite-difference approximations. The results are reported in Table \ref{tab:shocks}.
    We observe that the valuation adjustments remain stable around the baseline scenario, and that the corresponding finite-difference estimates of the sensitivities are themselves stable.

    These sensitivities are directional derivatives in a prescribed direction. A more holistic notion of sensitivity would be obtained by considering an upsilon ($\Upsilon$) sensitivity a la \cite{bartl2021sensitivity}, which we leave for future research.

    \begin{table}[H]
    \centering
    \begin{tabular}{|c|c|c|c|c|}
    \hline
    Shock $s$
    &  $\HVA^s_0$ & $\KVA^s_0$
    & $\dfrac{\HVA^s_0-\HVA^0_0}{s}$
    & $\dfrac{\KVA^s_0-\KVA^0_0}{s}$ \\\hline
    0           & 181.125 & 35.891 &              &              \\\hline
    0.00050     & 181.600 & 35.771 &   950.849    &  -239.752    \\\hline
    0.00025     & 181.363 & 35.831 &   951.344    &  -240.055    \\\hline
    -0.00050    & 180.648 & 36.011 &   952.829    &  -240.969    \\\hline
    -0.00025    & 180.887 & 35.951 &   952.334    &  -240.663    \\\hline
    \end{tabular}
    \caption{
    Time-0 $\HVA$ and $\KVA$ for parallel shocks $s$ on the jump intensity in the fair valuation model, together with finite-differences approximations of the corresponding sensitivities around the baseline scenario $s=0$.
    }
    \label{tab:shocks}
\end{table}

    \subsection{Not-So-Bad Trader}
    Regarding the not-so-bad trader, as $q^2(2,1)=0$, the option is called at $k=2$ if the model switch has not occurred before, hence all the $\Omega_{l,m},\;3 \leq l \leq 10,$ are equivalent to $\Omega_{11,11}$. %
    As for $l \le 2$, on  $\Omega_{l,m}$, the not-so-bad trader always calls the option at time $m$, which is the first time beyond $l$ for which $Q(m,I_m)=Q(m,1)=0$. Accordingly, we only report on the results corresponding to the events $\Omega_{l,m}$, for $l=1$ or $2$ and $m > l$, and $\Omega_{11,11}$.

    Figure \ref{fig:not-bad-hva-all} displays the not-so-bad trader's $\HVA$ (\textit{top left}) and its split in valuation (bottom) and early callability (\textit{top right}) components (see Proposition \ref{p:notsobad}(i)). Comparing with the bad trader's HVA components displayed in Figure \ref{fig:bad-hva}, we only see here $U^{nsb}$ and $va(K^{\texnsb})$ components, as the analogous processes $V^{nsb}$ and $W^{nsb}$ vanish as already observed in Remark \ref{rem: no W}. 
    The comparison with the top left panel of Figure \ref{fig:bad-hva} shows that
  $\HVA^{nsb}$ is more than twice smaller than $\HVA^{bad}$, but
    still significantly greater than the price difference $(q-Q)\ind_{[0,\ts)}$ 
    (%
    see the top left panel of Figure
    \ref{fig:not-bad-hva-all}%
    ).%
    
    Figures \ref{fig:not-bad-pnl} and \ref{fig:not-bad-hva-pnl}  display the not-so-bad trader's $\pal$ and HVA compensated pnl process.  
    As opposed to what we saw in Figure \ref{fig:bad-hva-pnl} regarding the bad trader, on the event $\Omega_{11,11}$, where there is no model switch and the not-so-bad trader calls back the claim according to the prescriptions of his wrong model, the gains resulting from the depreciation of the $\HVA$ no longer cover the $\pal$ losses (the dotted curve is in the positive in 
    Figure \ref{fig:not-bad-hva-pnl}): the better practice of switching to the global model once the trader's local model no longer calibrates not only diminishes the HVA, but also avoids the
    short-to-medium term incentives to use the local model. In fact, the local model does not pass the second
    Darwinian principle for the not-so-bad trader (see Subsection \ref{ss:da}), and would therefore not be selected by the latter (but only by the bad trader).

     \begin{figure}[H]
    \hspace*{-1cm} 
      \includegraphics[width=0.5\linewidth]{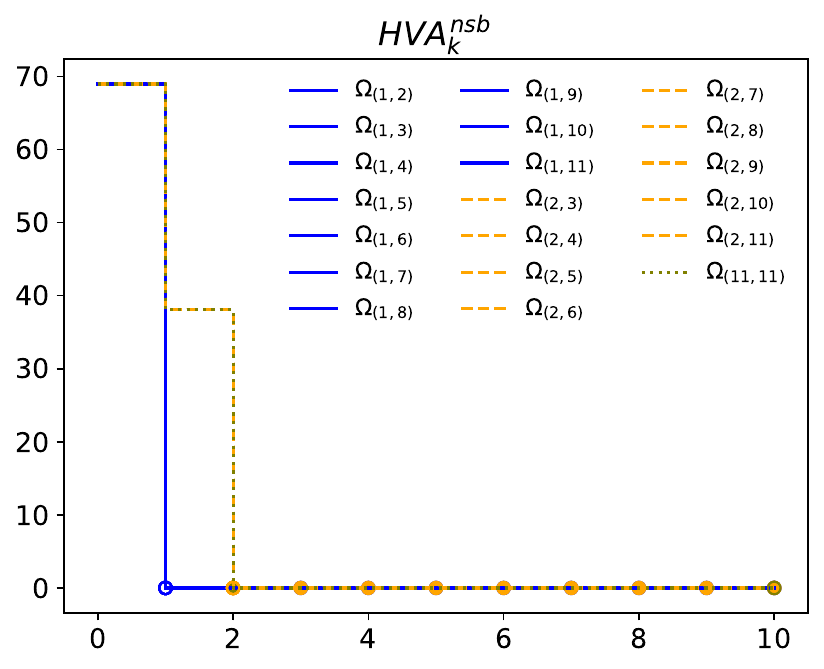} 
      \includegraphics[width=0.5\linewidth]{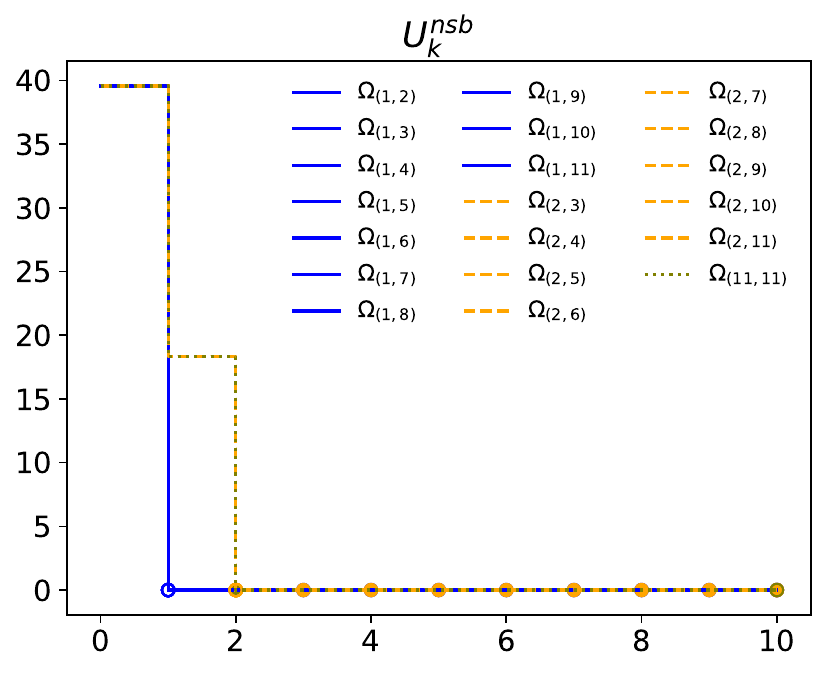}  \hspace{2cm}
      \centering
      \includegraphics[width=0.5\linewidth]{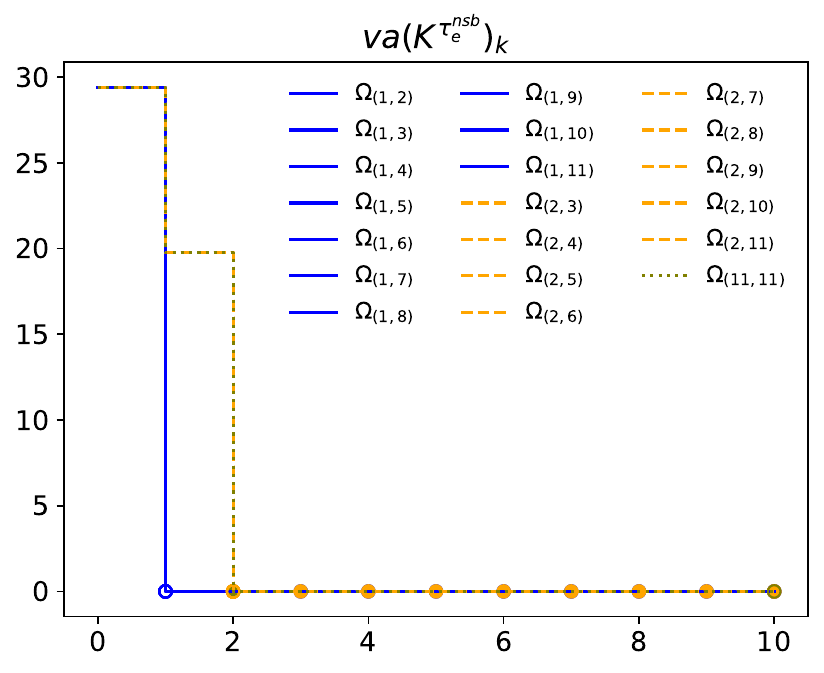} 
    \hspace*{-1cm}

    \caption{Not-so-bad trader's $\HVA$ and its components.}
    \label{fig:not-bad-hva-all}
    \end{figure}

    \begin{center}
    \begin{figure}[H] 
    \includegraphics[width=\linewidth]{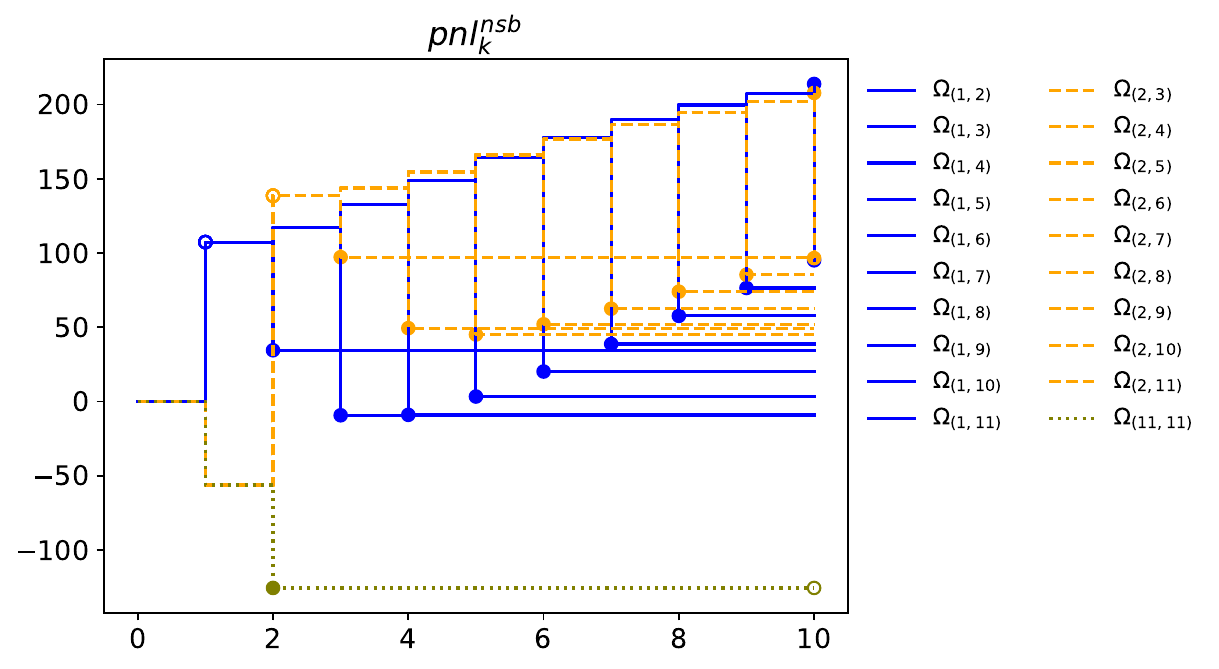}
    \caption{
    Not-so-bad trader's $pnl$.}
    \label{fig:not-bad-pnl}
    \end{figure}
    \end{center}
    
    \begin{figure}[H] 
     \includegraphics[scale=0.7]{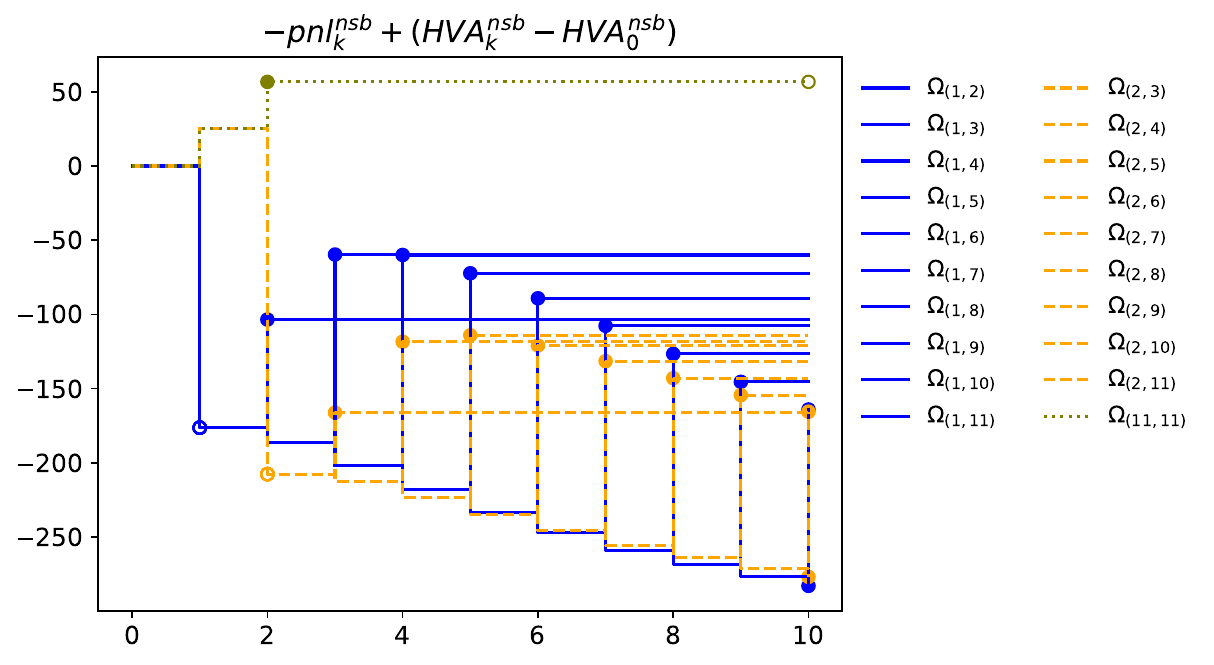}
    \caption{ 
    HVA-compensated loss-and-profits of the not-so-bad trader.}
    \label{fig:not-bad-hva-pnl}
    \end{figure}

    Figures \ref{fig:A-11} gathers on the same page the previous results for both traders in the event where the switch never happens, i.e.~on $\Omega_{11}$ in the case of the \bad trader and on $\Omega_{11,11}$ in the case of the not-so-bad one.
    The corresponding paths of the $\pal$ appear to be identical (as they indeed are) in the top panel of Figure \ref{fig:A-11}. As explained above, the $\HVA$ of the not-so-bad trader is smaller than the one of the bad one (\textit{middle panel});
    the HVA depreciation gains of the \bad trader fakely more than compensate %
    his raw $\pal$ losses (fakely in the sense that these systematic gains in fact only compensate future losses), but this is not the case for the not-so-bad trader (top and bottom panels), %
    
    \section{Conclusion\label{s:concl}}
    
    Figures \ref{fig:bad-es} and \ref{fig:not-bad-es} show the economic capital processes of the bad and not-so-bad traders of Propositions  \ref{p:bad}-\ref{p:notsobad}(iii), resulting in the KVA$_0$ displayed for a hurdle rate $h$ of 10\% in Table \ref{tab:HVA-KVA0}, along with the corresponding  HVA$_0$. 
    As expected, $\HVA^{nsb}_0 \leq \HVA^{bad}_0$ and $\KVA^{nsb}_0 \leq \KVA^{bad}_0$, \b{which illustrates
     the relevance of the proposed HVA and KVA metrics in terms of their sensitivities to the specification of the setup.} In this example, the 
    KVA is largely dominated by the HVA, by a factor $>4$, whereas the opposite was prevailing in the case of model risk on a European claim in \citet*[Eqn.~(37)]{AlbaneseCrepeyBenezet22}. However, a common and salient conclusion is that, in all the considered examples: bad or not-so-bad trader dealing a callable claim here or a European claim (for which bad or not-so-bad was in fact the same) in the previous paper, the risk-adjusted HVA, $\AVA=\HVA+\KVA$ (additional valuation adjustment for model risk), is much larger 
    than the price difference $q-Q$ of the claim between the trader's model and a reference model. 
    Whether this is mainly due to an HVA effect as in the present callable case (see Figures \ref{fig:bad-hva}  and  \ref{fig:not-bad-hva-all}) or to
    a KVA effect in \cite{AlbaneseCrepeyBenezet22}, in any case, it provides quantitative arguments in favour of a reserve for model risk that should be much larger than the common practice of reserving such a price difference simply (cf.~\citet*[Remark 2.10]{AlbaneseCrepeyBenezet22}).
    
This paper is focused on the Darwinian model risk of adverse selection by traders of local models motivated by short-to-medium gains at the expense of long term losses. We demonstrate how this can be a critical model risk issue regarding the handling of structured products by banks. This holds even disregarding the uncertainties, most commonly considered in the academic model risk literature and simply ignored for clarity in this work, regarding the risk-neutral and physical probability measures that underlie our fininsurance (global valuation) measure $\Q$ (see Subsection \ref{ss:se}).  We refer to \cite{bartl2021sensitivity}, specifically their Upsilon ($\Upsilon$) sensitivity, see also \cite{sauldubois2024first}, to assess quantitatively such uncertainties. 
In particular, our framework assumes access to a well-specified fair valuation model on which local models can be perfectly calibrated (or not anymore, at time $\tau_s$).
\b{Some insights into the sensitivity of our HVA and KVA metrics are provided in the paper by the consideration of the two traders, the bad and not-so-bad one, and the assessment of the impact of their different behaviour on the HVA and the KVA, as well as by the numerical stability study of Subsubection \ref{ss:numstab}. A more systematic investigation of HVA and KVA sensitivities, as well as the
incorporation of the uncertainty on the underlying physical and risk-neutral measures,
 are left for future research.

An important overarching question is: How far do we go in adding valuation adjustments? A distinguishing Darwinian model risk feature is that it cannot be detected by standard market risk metrics such as value-at-risk, expected shortfall or stressed value-at-risk. Indeed Darwinian model risk
derives from the  cumulative effect of daily recalibrations and feeds into the first moment of returns
(alpha leakages); the usual market risk metrics, instead, all focus on higher moments of return distributions at short-time horizons (such as one day).
Darwinian model risk can only be seen by simulating the hedging behavior of a bad model within the environment of good model. Even under the elementary probabilistic model under consideration in Sections \ref{s:dt} and \ref{s:num}, the computation of the valuation adjustments is nontrivial due to the nested recalibration of the local model at each node of the fair valuation one. So not only risk-adjusted reserve for model risk can be very high, but doing this for real portfolios and models, for which the (re)calibration can only be done numerically (as opposed to formulaically and exactly in our setup), would be far too demanding. We then do not advocate the banks to implement the HVA as an actual reserve on top of the regulatory requirements already implemented and which would affect their capital allocation strategies and lead to numerical and regulatory challenges. The primary aim of our work is not to propose immediate practical implementation, but rather to highlight an important risk: model risk inherent in using suboptimal or inadequate models is significantly greater than the pricing differences between good and bad models alone might suggest. Our methodology dissects this risk and makes explicit the hidden costs incurred.
But we rather view the HVA as a theoretical market discipline and a warning signal, encouraging the adoption of better model standards before practical implementation challenges arise.
The best practice would be that banks be more strongly incentized by regulators to only rely on high-quality models, so that such overwhelming computations (beyond tailor-made examples such as the one of this paper) are simply not needed.}

     \begin{figure}[H]
     \centering
    \includegraphics[scale=0.6]{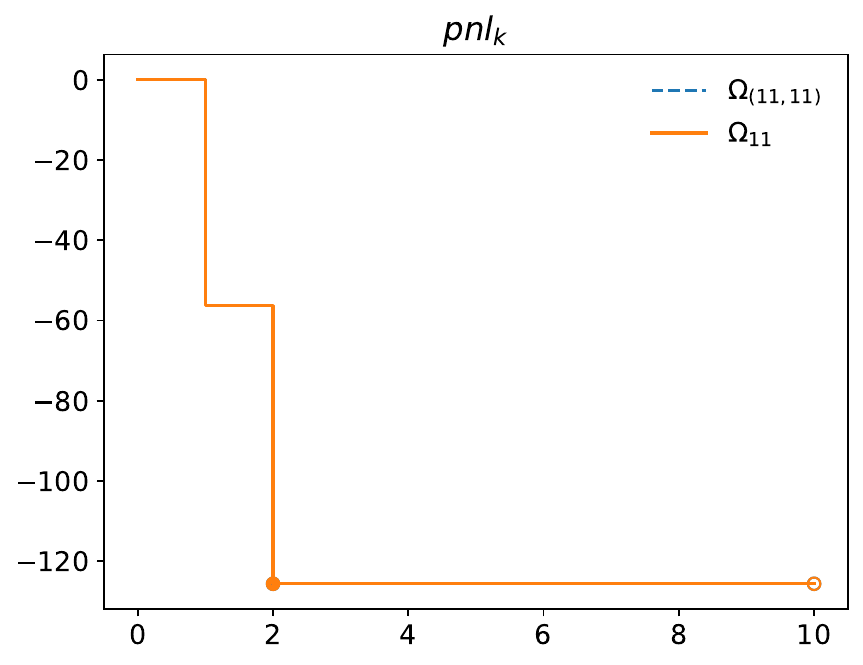}\\ 
    \includegraphics[scale=0.6]{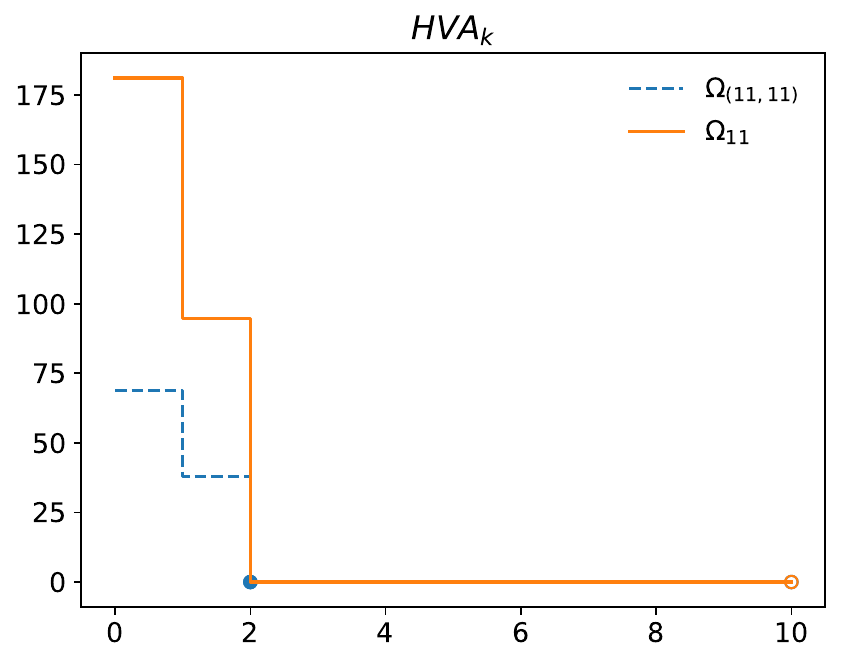}\\
    \includegraphics[scale=0.6]{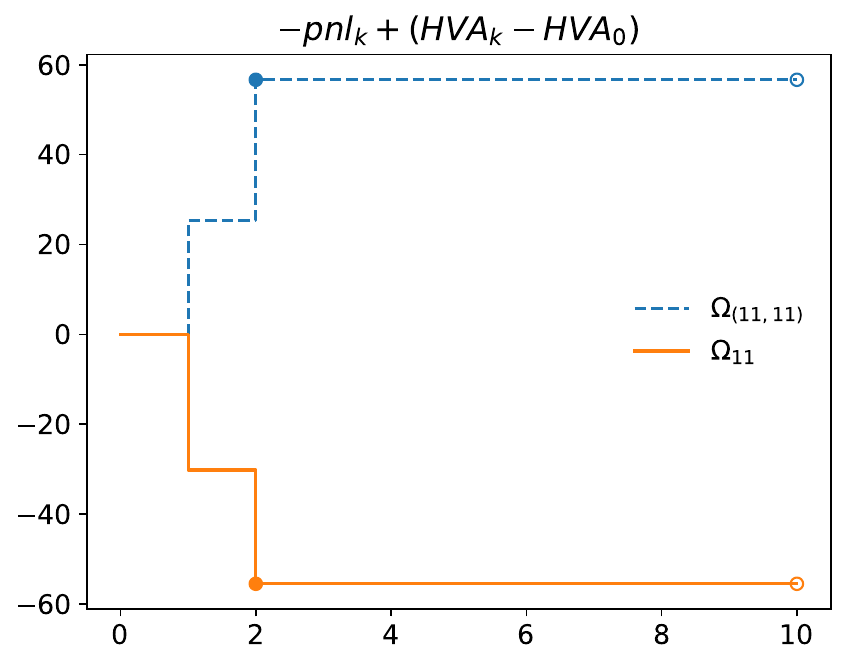}\\ 
    \caption{
    (\textit{top}) $pnl$, (\textit{center}) $\HVA$, and (\textit{bottom}) $\HVA_{(0)}-\pnl$ of the bad trader and the not-so-bad trader in the absence of model  switch. }
    \label{fig:A-11}
    \end{figure}

    \begin{figure}[H]
    \centering
    \includegraphics[scale=0.7]{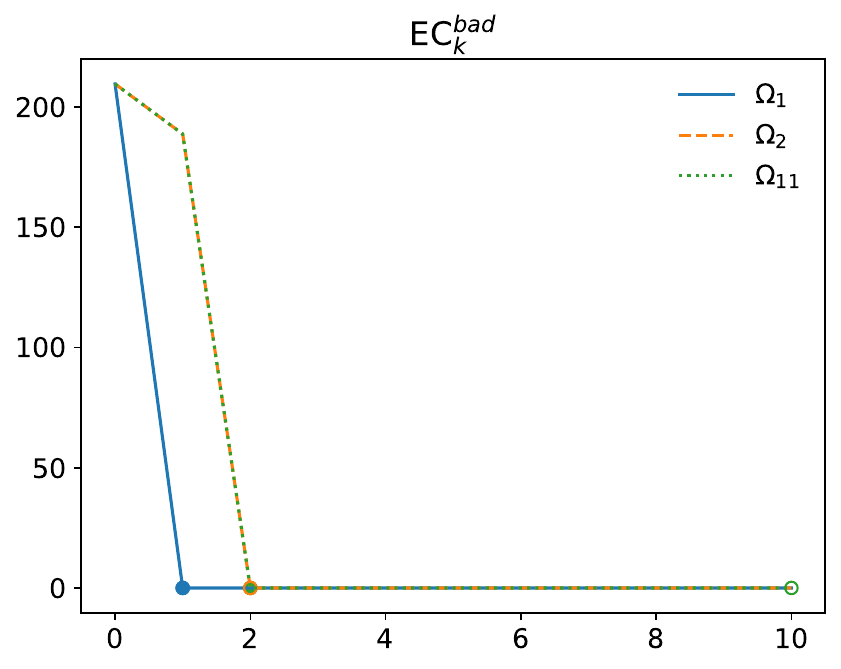}
    \caption{Bad trader's economic capital. 
    }
    \label{fig:bad-es}
    \end{figure}

    \begin{figure}[H]
    \centering
    \includegraphics[scale=0.7]{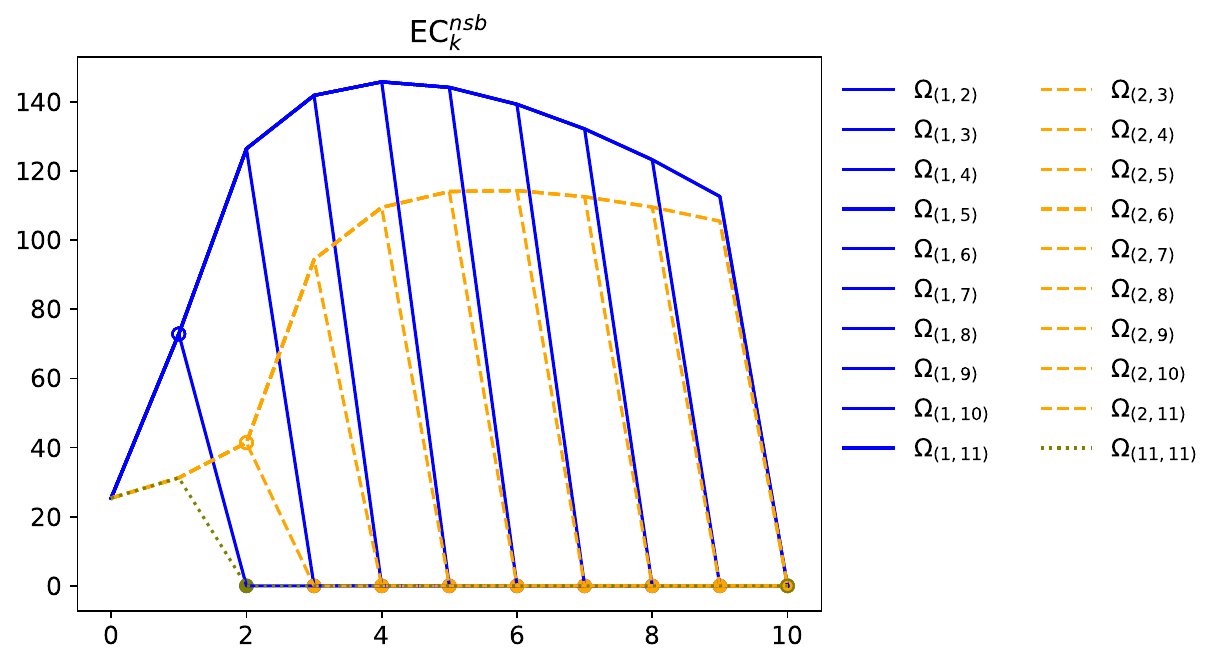}
    \caption{
    Not-so-bad trader's economic capital.}
    \label{fig:not-bad-es}
    \end{figure}

    \begin{table}[H]
    \centering
    \begin{tabular}{|c|c|c|}
        \hline
         &  $\HVA_0$ & $\KVA_0$\\\hline
     bad trader    & $181$ & $36$\\\hline
     not-so-bad trader    & $69$ & $15$\\\hline
    \end{tabular}
    \caption{
    $\HVA_0$ and $\KVA_0$ of the traders.}
    \label{tab:HVA-KVA0}
    \end{table}

    \appendix

    \section{Proofs of the Combinatorial Lemmas of Section \ref{s:dt}  \label{s:B}}
    \subsection{Bad Trader \label{ss:B1}}
    
    \paragraph{Proof of Lemma \ref{l:Rwo}} For $ 0<\lambda \le k$, $\Awo_\lambda$ is $\F_{\lambda}$ measurable, hence $\F_{k}$ measurable, thus $\probc{k}{\Awo_\lambda} = \ind_{\Awo_\lambda}$; in addition, for each $0 \le l \le\Td +1,$
    $$\ind_{\Awo_\lambda}(\Awo_l)=\un_{l=\lambda}.$$
    This proves
    $$
     \un_{k \ge \lambda>0}\probc{k}{\Awo_\lambda}(\Awo_l)= \un_{k \ge \lambda>0}\un_{l=\lambda}.
    $$
    
    Moreover, for each $0 \le k < \lambda \le\Td $, we compute
    \bel
    &\probc{k}{\Awo_{\Td +1}} =\prod_{m=1}^k \indi{N_m-N_{m-1} \mbox{ even}} \prod_{m=k+1}    \prob{N_m-N_{m-1} \mbox{ even}} \\&\qqq = \prod_{m=1}^k \indi{N_m-N_{m-1} \mbox{ even}} \prod_{m=k+1}    u_m ,\\
     & \probc{k}{\Awo_\lambda}= \big(\prod_{m=1}^{k} \indi{N^{}_m-N^{}_{m-1} \mbox{ even}} \big)\prob{N^{}_\lambda-N^{}_{\lambda-1} \mbox{ odd}}\times
     \prod_{m=k+1}^{\lambda-1} \prob{N^{}_m - N^{}_{m-1} \mbox{ even}} \\
     &\qqq =\prod_{m=1}^{k} \indi{N^{}_m-N^{}_{m-1} \mbox{ even}}
     \big(\prod_{m=k+1}^{\lambda-1}
     \even_{m } \big)\odd_{\lambda }    ,
    \eel 
    where, for each $0 \le l \le\Td +1,$
    $\big(\prod_{m=1}^{k} \indi{N^{}_m-N^{}_{m-1} \mbox{ even}}\big)(\Awo_l)=\un_{l>k}.$
    This proves
    $\un_{0 \le k < \lambda}\probc{k}{\Awo_\lambda}(\Awo_l)=  \un_{0 \le k < \lambda} \un_{l>k}\big(\prod_{m=k+1}^{\lambda-1} u_{m}  \big)v_\lambda   $
    as well as the last line in \eqref{r:rkll}. \finproof
    
 \paragraph{Proof of Lemma \ref{lem:exact wo}}    {\textbf{(i)}}  
    From \eqref{e:thetaudiscrs},
    $\texbad$ is a stopping time with respect to the filtration $\F^I$. Moreover, $\xi$ is measurable with respect to $\F^I_{\texbad}$,
    hence $\xi \indi{\texbad \le %
    l} $ is $\F^I_l$ measurable, for each $1 \le l \le\Td $. Therefore, for each $1 \le l \le\Td $, $\xi \indi{\texbad \le %
    l} = \Psi_l( I^{}_0, \dots, I^{}_l)$ holds for some map $\Psi_l:\{1,-1\}^{l+1} \to \mathbb R$.\\
    Note that $\texbad   \le \ts  = %
    l$ holds on $ \Awo_l$, i.e.~$\Awo_l \subseteq \{\texbad \le %
    l\}$.\\
    For $\omega \in \Awo_l$, we thus have $\xi(\omega) = \xi(\omega)\indi{\texbad(\omega) \le l} = \Psi_l(I^{}_0(\omega), \dots, I^{}_l(\omega)) = \Psi_l(1,\dots,1,-1)$, hence $\xi(\Awo_l)$ is well defined for $1 \le l \le\Td $.\\
    Similarly, for all $\omega \in \Awo_{\Td +1}$, one has $\xi(\omega) = \Psi_T(1,\dots,1)$, hence $\xi(\Awo_{\Td +1})$ is also well defined.
    {\rm\hfill\break \textbf{(ii)}} Since $\xi$ is constant on each $\Awo_\lambda$, $1 \le \lambda \le T+1$, which partition $\Omega$, the $\F_k$ conditional law of $\xi$ is given, for all $0 \le k \le T$, by
    \beql{e:clk}
    \cL_k(\xi) = \sum_{\lambda=1}^{T+1} \R_k[\Awo_\lambda] \boldsymbol\delta_{\xi(\Awo_\lambda)}.
    \eeql
    By Lemma \ref{l:Rwo}, $\R_k[\Awo_\lambda]$ is constant on each $\Awo_l$, implying that $\cL_k(\xi)$ is also constant on each $\Awo_l$, $1 \le l \le T$. In particular, $\espc{k}{\xi}, \VaR_k(\xi)$ and $\ES_k(\xi)$ are constant on each $\Awo_l$. We also compute
    \bel 
    &\espc{k}{\xi}
     = \sum_{\lambda=1}^{\Td+1}\xi(\Awo_\lambda)\probc{k}{\Awo_\lambda} = \sum_l \ind_{\Awo_l}\sum_{\lambda=1}^{\Td+1}
    \xi(\Awo_\lambda) \probc{k}{\Awo_\lambda}(\Awo_l).
    \eel
Last, if $l \le k$, then Lemma \ref{l:Rwo} yields $\probc{k}{\Awo_\lambda}(\Awo_l) = \mathbf{1}_{\lambda=l}$, hence \eqref{e:clk} reduces to $$\cL_k(\xi)(\Omega_l) = \R_k[\Awo_l](\Awo_l) \boldsymbol\delta_{\xi(\Awo_l)} = \boldsymbol\delta_{\xi(\Awo_l)},$$ which implies that $\espc{k}{\xi}(\Awo_l)=\VaR_k[\xi](\Awo_l)=\ES_k[\xi](\Awo_l)=\xi(\Awo_l)$.
     \finproof
     
    \subsection{Not-So-Bad Trader \label{ss:B2}}
    
\paragraph{Proof of Lemma \ref{l:rkllno}}  For each $(l,m)\in\cI$,
    all paths of $I$ represented in $\Ano_{l,m}$ have the same beginning until time step $m\wedge\Td $.
    We denote by $\Ano_{l,m}^k$ 
    the event defined by this beginning of the path of $I$ until time step $k\le m\wedge\Td $.  
    
    We compute
    \bel
    & \probc{k}{\Ano_{\Td +1,\Td +1}} = \prod_{r=1}^{\Td} \probc{k}{N^{}_r-N^{}_{r-1} \mbox{ even}}= \\ & \qqq\prod_{r=1}^k \indi{N_r-N_{r-1} \mbox{ even}} \prod_{r=k+1}^{\Td}   \prob{N_r-N_{r-1} \mbox{ even}} =\ind_{\Ano_{\Td +1,\Td +1}^k}\prod_{r=k+1}^{\Td}   u_r, 
     \eel
     where $
    \ind_{\Ano_{\Td +1,\Td +1}^k}=\sum_{1\le l\le m\le\Td }\ind_{\Ano_{l,m}}\un_{k\le l} +\sum_{l
    \le\Td }\ind_{\Ano_{l,\Td +1}}\un_{k< l} + \ind_{\Ano_{\Td +1,\Td +1}}  , $
     which proves the last identity in \eqref{r:rkllno}.
    
    Similarly, for $1\le\lambda\le\Td ,$
    \bel 
      \probc{k}{\Ano_{\lambda,\Td +1}} =& 
    \ind_{\Ano_{\lambda,\Td +1}^k}
      \times 
    \Big(\un_{ k\ge\lambda } 
     \prod_{r=k+1 }^{\Td } \even_{r}+\un_{ k<\lambda } 
    \big( \prod_{r=k+1 }^{\lambda-1 } \even_{r}\big)
     \odd_{\lambda}
     \big(\prod_{r=\lambda+1 }^{\Td } \even_{r}\big)\Big),    
      \eel
      where
      \bel
    &\ind_{\Ano_{\lambda,\Td +1}^k}(\Og_{l,m})= 
    \un_{1\le l\le m\le\Td }
    (\un_{k<l\wedge \lambda }+\un_{k\ge  l\wedge \lambda }\un_{l= \lambda }\un_{ k<m })
    +\\&\qqq\qqq
    \un_{1\le l\le \Td ,m=\Td +1}
    (\un_{k<l\wedge \lambda }+\un_{k\ge  l\wedge \lambda }\un_{l= \lambda } )
    + \un_{l=\Td +1,m=\Td +1}\un_{k<l\wedge \lambda }\\&
     \qqq=\un_{k<l\wedge \lambda }+\un_{k\ge  l\wedge \lambda }\un_{l= \lambda }\un_{ k<m }, 
     \eel
    which proves the second identity in \eqref{r:rkllno}.
    
    Finally, for $1\le\lambda\le \mu\le\Td ,$
    \bel
     & \probc{k}{\Ano_{\lambda,\mu }}(\Ano_{l,m})
    = \ind_{\Ano_{\lambda,\mu}^k}
    \times\\&\qqq
    \Big(\un_{ k\ge\mu }+
    \un_{\lambda\le  k<\mu } 
    \big( \prod_{r=k+1 }^{\mu-1 } \even_r\big)
     \odd_{\mu}  + \un_{ k<\lambda } 
    \big( \prod_{r=k+1 }^{\lambda-1 } \even_{r}\big)
     \odd_{\lambda}
     \big(\prod_{r=\lambda+1 }^{\mu-1} \even_{r}\big) \odd_{\mu }\Big) ,    \eel
    where 
      \bel
    &\ind_{\Ano_{\lambda,\mu}^k}(\Og_{l,m})= 
    \un_{1\le l\le m\le\Td }
    \big(\un_{k<l\wedge \lambda }+\un_{k\ge  l\wedge \lambda }\un_{l= \lambda }(\un_{ k<m\wedge \mu }+\un_{ k\ge m\wedge \mu }\un_{m=\mu })\big)
    +\\&\qqq\qqq
    \un_{1\le l\le \Td ,m=\Td +1}
    (\un_{k<l\wedge \lambda }+\un_{k\ge  l\wedge \lambda }\un_{l= \lambda } \un_{ k<\mu   } )
    + \un_{l=\Td +1,m=\Td +1}\un_{k< \lambda }\\&\qqq=
     \un_{k<l\wedge \lambda }+\un_{k\ge  l\wedge \lambda }\un_{l= \lambda }(\un_{ k<m\wedge \mu }+\un_{ k\ge m\wedge \mu }\un_{m=\mu }),\eel which proves the first identity in \eqref{r:rkllno}. \finproof

\paragraph{Proof of Lemma \ref{lem:exact no}}   {\textbf{(i)}}
    Since $\texnsb$ is an $\F^I$ stopping time and 
    $\xi$ is $\F^I_{\texnsb}$ measurable,
    it follows that 
    $\xi \indi{\texnsb \le  m} $ is 
    $\F^I_m$ measurable, for each $1 \le m \le \Td $. 
    We thus have, for all $1 \le m \le\Td $, $\xi\indi{\texnsb \le %
    m} = \Psi_{m}( I^{}_0, \dots, I^{}_m)$ for some map $\Psi_{m}:\{1,-1\}^{m+1} \to \mathbb R$.\\
    For $\omega \in \Awo_{l,m}$ such that $1 \le l < m\le\Td $, we have 
    $\tex(\omega) \le m$, i.e.~$\Awo_{l,m} \subseteq \{\texnsb \le %
    m\}$, and hence $
    \xi(\omega) = \xi(\omega)\indi{\texnsb(\omega) \le %
    m} = \Psi_{m}(I^{}_0(\omega), \dots, I^{}_m(\omega)) = \Psi_m(-1,\dots,-1,1,\dots,1,-1)$,
    hence $\xi(\Awo_{l,m})$ is well defined for $1 \le l \le m\le\Td $.\\
    Moreover $I$ and therefore $\xi$ are constant on each
      $ \Ano_{l,\Td +1}$ such that $1 \le l  \le\Td +1$, hence   $\xi(\Ano_{l,\Td +1})$ is also well defined for each $1 \le l  \le\Td +1$.
    {\rm\hfill\break \textbf{(ii)}} Since $\xi$ is constant on each $\Ano_{\lambda,\nu}$, $(\lambda,\nu)\in\cI$, which partition $\Omega$, the $\F_k$ conditional law of $\xi$ is given, for all $0 \le k \le T$, by
    \bel
    \cL_k(\xi) = \sum_{(\lambda,\nu)\in\cI} \R_k[\Omega_{\lambda,\nu}] \boldsymbol\delta_{\xi(\Omega_{\lambda,\nu})}.
    \eel
    By Lemma \ref{l:rkllno}, $\R_k[\Ano_{\lambda,\nu}]$ is constant on each $\Ano_{l,m}$, implying that $\cL_k(\xi)$ is also constant on each $\Awo_{l,m}$, $(l,m)\in\cI$. In particular, $\espc{k}{\xi}, \VaR_k(\xi)$ and $\ES_k(\xi)$ are constant on each $\Ano_{l,m}$. Last, we compute
    \bel 
    &\espc{k}{\xi}
     = \sum_{(\lambda,\nu)\in\cI}\xi(\Ano_{\lambda,\nu})\probc{k}{\Ano_{\lambda,\nu}} = \sum_{(l,m)\in\cI} \ind_{\Ano_{l,m}}\sum_{(\lambda,\nu)\in\cI}
    \xi(\Ano_{\lambda,\nu}) \probc{k}{\Ano_{\lambda,\nu}}(\Awo_{l,m}). \,
    \finproof
    \eel

\bibliographystyle{chicago}
\bibliography{ref}

\begin{thebibliography}{}

\bibitem[\protect\citeauthoryear{Albanese, Cr\'epey, Hoskinson, and Saadeddine}{Albanese et~al.}{2021}]{CrepeyHoskinsonSaadeddine2019}
Albanese, C., S.~Cr\'epey, R.~Hoskinson, and B.~Saadeddine (2021).
\newblock {XVA} analysis from the balance sheet.
\newblock {\em Quantitative Finance\/}~{\em 21\/}(1), 99--123.

\bibitem[\protect\citeauthoryear{Albanese, Cr{\'e}pey, and Iabichino}{Albanese et~al.}{2021}]{AlbaneseCrepeyIabichino19}
Albanese, C., S.~Cr{\'e}pey, and S.~Iabichino (2021).
\newblock A {D}arwinian theory of model risk.
\newblock {\em Risk Magazine\/}.
\newblock July, 72--77.

\bibitem[\protect\citeauthoryear{Artzner, Eisele, and Schmidt}{Artzner et~al.}{2024}]{ArtznerEiseleSchmidt22}
Artzner, P., K.-T. Eisele, and T.~Schmidt (2024).
\newblock Insurance--finance arbitrage.
\newblock {\em Mathematical Finance\/}~{\em 34\/}(3), 739--773.

\bibitem[\protect\citeauthoryear{Barrieu and Scandolo}{Barrieu and Scandolo}{2015}]{barrieu2015assessing}
Barrieu, P. and G.~Scandolo (2015).
\newblock Assessing financial model risk.
\newblock {\em European Journal of Operational Research\/}~{\em 242\/}(2), 546--556.

\bibitem[\protect\citeauthoryear{Bartl, Drapeau, Ob{\l}{\'o}j, and Wiesel}{Bartl et~al.}{2021}]{bartl2021sensitivity}
Bartl, D., S.~Drapeau, J.~Ob{\l}{\'o}j, and J.~Wiesel (2021).
\newblock Sensitivity analysis of {W}asserstein distributionally robust optimization problems.
\newblock {\em Proceedings of the Royal Society A\/}~{\em 477\/}(2256), 20210176.

\bibitem[\protect\citeauthoryear{B{\'e}n{\'e}zet and Cr{\'e}pey}{B{\'e}n{\'e}zet and Cr{\'e}pey}{2024}]{AlbaneseCrepeyBenezet22}
B{\'e}n{\'e}zet, C. and S.~Cr{\'e}pey (2024).
\newblock Handling model risk with {XVA}s.
\newblock {\em Frontiers of Mathematical Finance\/}~{\em 3\/}(4), 490--519.

\bibitem[\protect\citeauthoryear{Burnett}{Burnett}{2021}]{Burnett21}
Burnett, B. (2021).
\newblock Hedging value adjustment: Fact and fiction.
\newblock {\em Risk Magazine\/}.
\newblock February, 1--6.

\bibitem[\protect\citeauthoryear{Burnett, McCrickerd, and Piau}{Burnett et~al.}{2025}]{burnett2025fundamental}
Burnett, B., R.~McCrickerd, and B.~Piau (2025).
\newblock The fundamental representation of pricing adjustments.
\newblock arXiv:2503.14997.

\bibitem[\protect\citeauthoryear{Burnett and Williams}{Burnett and Williams}{2021}]{Burnett21b}
Burnett, B. and I.~Williams (2021).
\newblock The cost of hedging {XVA}.
\newblock {\em Risk Magazine\/}.
\newblock April.

\bibitem[\protect\citeauthoryear{Cr\'epey}{Cr\'epey}{2022}]{Crepey21}
Cr\'epey, S. (2022).
\newblock Positive {XVA}s.
\newblock {\em Frontiers of Mathematical Finance\/}~{\em 1\/}(3), 425--465.

\bibitem[\protect\citeauthoryear{Cr\'epey}{Cr\'epey}{2025}]{Crepey25}
Cr\'epey, S. (2025).
\newblock {\em {XVA Analysis: Probabilistic, Risk Measure, and Machine Learning Issues}}.
\newblock Taylor \& Francis, New York.
\newblock Chapman \& Hall/CRC Financial Mathematics Series.

\bibitem[\protect\citeauthoryear{El~Karoui}{El~Karoui}{1981}]{ElKaroui1981SaintFlour}
El~Karoui, N. (1981).
\newblock Les aspects probabilistes du contr{\^o}le stochastique.
\newblock In P.~L. Hennequin (Ed.), {\em Ecole d'Et{\'e} de Probabilit{\'e}s de Saint-Flour IX-1979}, pp.\  73--238. Berlin, Heidelberg: Springer.

\bibitem[\protect\citeauthoryear{Fan, Park, and Xu}{Fan et~al.}{2025}]{fan2025quantifying}
Fan, Y., H.~Park, and G.~Xu (2025).
\newblock Quantifying distributional model risk in marginal problems via optimal transport.
\newblock {\em Mathematics of Operations Research\/}.

\bibitem[\protect\citeauthoryear{Gianfreda and Scandolo}{Gianfreda and Scandolo}{2024}]{gianfreda2024assessing}
Gianfreda, A. and G.~Scandolo (2024).
\newblock Assessing model risk in financial and energy markets using dynamic conditional vars.
\newblock {\em Applied Stochastic Models in Business and Industry\/}~{\em 40\/}(2), 408--433.

\bibitem[\protect\citeauthoryear{Lazar, Qi, and Tunaru}{Lazar et~al.}{2024}]{lazar2024measures}
Lazar, E., S.~Qi, and R.~Tunaru (2024).
\newblock Measures of model risk for continuous-time finance models.
\newblock {\em Journal of Financial Econometrics\/}~{\em 22\/}(5), 1456--1481.

\bibitem[\protect\citeauthoryear{Matsumoto and Suyama}{Matsumoto and Suyama}{2024}]{matsumoto2024multi}
Matsumoto, K. and T.~Suyama (2024).
\newblock Multi-period mean--variance hedging problem with model risk.
\newblock {\em Applied Mathematical Finance\/}~{\em 31\/}(6), 365--384.

\bibitem[\protect\citeauthoryear{Neveu}{Neveu}{1975}]{neveu1975discrete}
Neveu, J. (1975).
\newblock {\em Discrete-Parameter Martingales}.
\newblock Mathematical Studies. North-Holland.

\bibitem[\protect\citeauthoryear{Sauldubois and Touzi}{Sauldubois and Touzi}{2024}]{sauldubois2024first}
Sauldubois, N. and N.~Touzi (2024).
\newblock First order martingale model risk and semi-static hedging.
\newblock arXiv:2410.06906.

\bibitem[\protect\citeauthoryear{Silotto, Scaringi, and Bianchetti}{Silotto et~al.}{2024}]{silotto2024xva}
Silotto, L., M.~Scaringi, and M.~Bianchetti (2024).
\newblock Xva modelling: validation, performance and model risk management.
\newblock {\em Annals of Operations Research\/}~{\em 336\/}(1), 183--274.

\end{thebibliography}

\end{document}